\documentclass[journal=jctcce,manuscript=article]{achemso}
\setkeys{acs}{articletitle = true}
\usepackage{chemformula}
\usepackage[T1]{fontenc}
\usepackage{blindtext}
\usepackage{amsfonts}
\usepackage{enumitem}
\usepackage{amssymb}
\usepackage{multirow}
\usepackage{xcolor}
\usepackage[ruled,vlined]{algorithm2e}

\newcommand{\onlinecite}[1]{\hspace{-1 ex} \nocite{#1}\citenum{#1}}
\def\half{\frac{1}{2}}

\usepackage{accents}

\newcommand{\nsvd}{N_{\mathrm{SVD}}}
\newcommand{\nqua}{N_{\mathrm{qua}}}

\DeclareUnicodeCharacter{2212}{-}

\author{Micha{\l} Lesiuk}
\email{m.lesiuk@uw.edu.pl}
\affiliation{Faculty of Chemistry, University of Warsaw\\ Pasteura 1, 02-093 Warsaw, Poland}
\date{\today}

\title[]{When gold is not enough: platinum standard of quantum chemistry with $N^7$ cost}

\keywords{coupled-cluster theory, tensor decomposition}

\begin{document}

\begin{tocentry}
\includegraphics[scale=0.82]{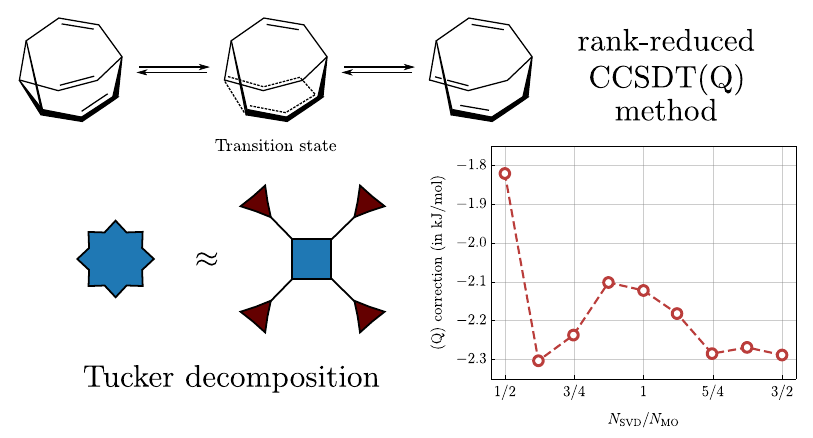}
\end{tocentry}

\renewcommand{\arraystretch}{1.5}

\begin{abstract}
In this paper we extend the rank-reduced coupled-cluster formalism to the calculation of non-iterative energy corrections due to quadruple excitations. There are two major components of the proposed formalism. The first is an approximate compression of the quadruple excitation amplitudes using the Tucker format. The second is a modified functional used for evaluation of the corrections which gives exactly the same results for the exact amplitudes, but is less susceptible to errors resulting from the aforementioned compression. We show, both theoretically and numerically, that the computational cost of the proposed method scales as the seventh power of the system size. Using reference results for a set of small molecules, the method is calibrated to deliver relative accuracy of a few percent in energy corrections. To illustrate the potential of the theory we calculate the isomerization energy of \emph{ortho}/\emph{meta} benzyne (C$_6$H$_4$) and the barrier height for the Cope rearrangement in bullvalene (C$_{10}$H$_{10}$). The method retains a near-black-box nature of the conventional coupled-cluster formalism and depends on only one additional parameter that controls the accuracy. 
\end{abstract}

\newpage
\section{Introduction}
\label{sec:intro}

Tensor decomposition has long been an active area of research in the field of applied mathematics, 
with successful applications in many branches of science, see Ref.~\onlinecite{kolda09} for an exhaustive review. In 
recent years, tensor decomposition techniques have been embraced by the quantum chemistry 
community, as exemplified by the development of the tensor hypercontraction (THC) format of 
the electron repulsion integrals~\cite{hohenstein12,parrish12,parrish13a,parrish13b}. Pioneering applications to electronic structure methods such as MP2, MP3, random-phase approximation and coupled cluster have also been reported~\cite{kinoshita03,benedikt13,hohenstein13a,shenvi14,schumacher15,lesiuk19,parrish19,lee20,lesiuk20,matthews21,lesiuk21,lesiuk22}. The primary 
motivation for applying tensor decompositions to quantum-chemical methods are reductions in terms 
of computational cost and storage requirements. With a proper calibration these benefits are 
attainable with an insignificant accuracy loss and without compromising the black-box nature of the 
parent theoretical method. However, we would like to point out that from the point of view of quantum chemistry, there is an additional potential application of tensor decomposition techniques which has been largely untapped thus far. It is related to the interpretative power of such techniques, exploiting the fact that they can automatically extract important information about the system even from a complicated wavefunction Ans\"{a}tz,  with minimal human oversight~\cite{kolda09}.

The coupled-cluster (CC) theory~\cite{crawford07,bartlett07} is a particularly promising candidate for applying tensor 
decomposition schemes. In all CC variants the wavefunction is parametrized by a set of cluster 
amplitudes which can be viewed as multi-dimensional tensors with indices referring to the occupied 
and virtual orbital sets. Storage and manipulation of these tensors constitutes the main bottleneck 
in CC calculations for large molecular systems. To address this issue we have recently introduced~\cite{lesiuk20} 
an approximate CC theory including single, double and triple excitations (CCSDT)~\cite{noga87,scuseria88} where the 
triply-excited amplitudes tensor is represented in the Tucker-3 format~\cite{tucker66,delath00}. The increased flexibility 
offered by this decomposition enables to reduce the scaling of the approximate method by a factor 
quadratic in the system size in comparison with the exact CCSDT. At the same, accuracy levels of up 
to 0.1 kJ/mol are reachable in typical applications to chemical problems.

Despite these developments are promising, one may argue from a pragmatic standpoint that being able 
to reproduce the CCSDT results accurately, even at a significantly reduced cost, is not sufficient 
for general-purpose applications in thermochemistry, chemical kinetics, molecular interactions, etc. 
In fact, it is well-documented that in some applications the CCSDT method does not improve the 
accuracy (in relation to FCI) over the ``gold standard'' CCSD(T) to a degree that would justify the drastic increase in 
the computational costs~\cite{gauss02,rezac13,simonova13,smith14,karton19,kodrycka19}. The reason for this counterintuitive behavior is an accidental, yet 
systematic, cancellation of errors observed at the CCSD(T) level of theory for ``well-behaved'' 
systems. There are two major components of the post-CCSD(T) contribution: (i) the correction to 
due the inexact treatment of triple excitations and (ii) the correction due to the missing 
quadruple excitations. It turns out that these two components are often of opposite signs and 
hence a degree of cancellation occurs. The lesson learned is that the quadruple 
excitations play a significant role in the~$\approx1\,$kJ/mol accuracy range and must be included 
alongside the full treatment of triple excitations to provide a balanced description.

The importance of quadruple excitations in accurate theoretical studies was recognized in 
the literature a long time ago. Unfortunately, their complete inclusion by means of the full CCSDTQ 
theory~\cite{kucharski91,oliphant91,kucharski92,kucharski10} is prohibitive for molecules comprising more than a few atoms, assuming a decent-quality
basis set is used. This prompted research into more affordable methods that are able to account for 
the quadruple excitations in an approximate, yet still reliable, way. Several families of such 
methods were proposed~\cite{bartlett90,kucharski98,kucharski98b,bomble05,kallay05,kallay08,eriksen14a,eriksen14b}, both iterative and non-iterative, based either on the ordinary 
M\o{}ller-Plesset perturbation theory or various effective Hamiltonian approaches, and employing 
either CCSD or CCSDT wavefunctions as the starting point. A more detailed technical discussion of 
these methods is given in subsequent sections. In this work we concentrate primarily on the 
CCSDT(Q) theory introduced by Bomble \emph{et al}.~\cite{bomble05} which has become \emph{de facto} standard in 
high-accuracy quantum chemical calculations. Due to a good balance between the accuracy and 
computational costs, it is a member of various composite electronic structure protocols and is 
implemented is several program packages available for public use. In many applications, the 
CCSDT(Q) theory is considered to be the ``platinum standard'' of quantum chemistry~\cite{kodrycka19} -- the next 
rung of the coupled-cluster ladder above CCSD(T) striking a balance between the accuracy and computational costs. Computation of the (Q) correction is usually $1-2$ orders of magnitude less computationally 
intensive than the complete CCSDTQ calculations. Despite this drastic reduction, the range of 
applicability of the CCSDT(Q) theory to polyatomic molecules remains limited as a result of steep 
$N^9$ scaling of the computational costs with the system size, $N$. 

This work is a continuation of a series of papers~\cite{lesiuk19,lesiuk20,lesiuk21,lesiuk22} where tensor decomposition techniques are applied as a tool to reduce the cost of high-order CC methods. In this part, we introduce a rank-reduced approach to computation of the (Q) correction. There are two main distinguishing features of the proposed scheme. The first is the compression of the quadruply-excited amplitudes using the Tucker format
which enables to reduce the immense cost of storing and manipulating the $T_4$ amplitudes. To achieve 
the necessary transformation from the full rank to rank-reduced representation of the 
quadruply-excited amplitudes we develop an iterative method based on higher-order orthogonal 
iteration (HOOI) procedure~\cite{delath00b,elden09}. The second feature is the development of a modified functional used to evaluate the (Q) correction. Due to the variational nature of this functional, it is less sensitive to the errors incurred by the rank-reduced treatment of the CC amplitudes. This enables to evaluate the (Q) correction with a mean relative accuracy of a few percent. Taking into account that the (Q) method itself is able to recover, on average, about 90\% of the CCSDTQ-CCSDT energy difference~\cite{eriksen15}, these errors are acceptable from a practical point of view. Critically, by properly factorizing the working expression of the proposed method and exploiting the rank-reduced format of the CC amplitudes, it is possible to evaluate the (Q) correction with the $N^7$ cost. Finally, we report calculations of relative energies for larger systems, demonstrating a broad range of applicability and reliability of the proposed theory. In particular, we study the isomerization energy of \emph{ortho}/\emph{meta} benzyne and the Cope rearrangement in bullvalene molecule.

\section{Theory}
\label{sec:theory}

\subsection{Preliminaries}
\label{subsec:pre}

In this work we consider closed-shell systems and employ the canonical restricted Hartree-Fock (HF) determinant, denoted $|\phi_0\rangle$, as the reference wavefunction in the CC theory. The HF orbital energies are denoted by $\epsilon_p$. For brevity, we also introduce the following conventions: $\langle A\rangle \stackrel{\mbox{\tiny def}}{=} \langle \phi_0 | A \phi_0 \rangle$ and $\langle  A|B\rangle \stackrel{\mbox{\tiny def}}{=} \langle A \phi_0|B \phi_0 \rangle$ for arbitrary operators $A$, $B$. Unless explicitly stated otherwise, the Einstein convention for summation over repeated indices is employed throughout. The standard partitioning of the electronic Hamiltonian, $H=F+W$, into the sum of  the Fock operator ($F$) and the fluctuation potential ($W$) is adopted. The remaining aspects of the notation are summarized in  Table~\ref{tab:notation}.

\begin{table}[t]
  \caption{Details of the notation adopted in the present work; $O$ is the number of active
  occupied in the reference, $V$ is the number of virtual orbitals. For convenience of the readers 
the  key defining equations were included in the last column.}
  \label{tab:notation}
  \begin{tabular}{lclc}
    \hline
    Indices & Limit & 
    \multicolumn{1}{c}{Corresponds to} & 
    \multicolumn{1}{c}{Defining equation} \\
    \hline
    $i$, $j$, $k$, $l,\ldots$ & $O$ & 
    active occupied orbitals  & 
    \multicolumn{1}{c}{--} \\
    $a$, $b$, $c$, $d,\ldots$ & $V$ & 
    unoccupied (virtual) orbitals & 
    \multicolumn{1}{c}{--} \\
    $p$, $q$, $r$, $s,\ldots$ & 
    \multicolumn{1}{c}{--} &
    general orbitals          & 
    \multicolumn{1}{c}{--}  \\
    $P$, $Q,\ldots$           & $N_{\mathrm{aux}}$ &
    density-fitting basis set & 
    $(pq|rs) = B_{pq}^Q\,B_{rs}^Q$ \\
    $X$, $Y$, $Z,\ldots$      & $N_{\mathrm{SVD}}$ &
    subspace of triply-excited amplitudes     & 
    $t_{ijk}^{abc} = t_{XYZ} \,U^X_{ai}\,U^Y_{bj} \,U^Z_{ck}$ \\
    $A$, $B$, $C,\ldots$      & $N_{\mathrm{qua}}$ &
    subspace of quadruply-excited amplitudes  & 
    $t_{ijkl}^{abcd} = t_{ABCD} \,V^A_{ai}\,V^B_{bj}\,V^C_{ck}\,V^D_{dl}$ \\
    \hline
  \end{tabular}
\end{table}

The method for evaluation of non-iterative quadruples correction reported in this work builds 
upon the SVD-CCSDT theory introduced in Ref.~\onlinecite{lesiuk20}. The electronic wavefunction 
underlying the SVD-CCSDT method is given by $|\Psi\rangle = e^{T_{\rm SVD}}\,|\phi_0\rangle$ with 
$T_{\rm 
SVD}=T_1+T_2+T_3^{\rm SVD}$. The $T_1$ and $T_2$ operators have the same form as in the usual CCSDT 
theory
\begin{align}
\label{t12}
 T_1 = t_i^a\,E_{ai}, \;\;\;
 T_2 = \frac{1}{2}\,t_{ij}^{ab} \,E_{ai}\,E_{bj},
\end{align}
where $t_i^a$, $t_{ij}^{ab}$ are the cluster amplitudes, and $E_{pq}=p^\dagger_\alpha q_\alpha + 
p^\dagger_\beta q_\beta$ are the spin-adapted singlet orbital replacement 
operators~\cite{paldus88}. The triply excited component of the cluster operator is approximated as
\begin{align}
\label{t3}
 T_3^{\rm SVD} = \frac{1}{6}\, t_{ijk}^{abc}\,E_{ai}\,E_{bj}\,E_{ck},
 \;\;\;\mbox{with}\;\;\;
 t_{ijk}^{abc} \approx t_{XYZ} \,U^X_{ai}\,U^Y_{bj} \,U^Z_{ck}.
\end{align}
The quantities $U^X_{ai}$ are obtained by a procedure described in Ref.~\onlinecite{lesiuk22} and are fixed during the coupled-cluster iterations. The remaining unknown 
quantities ($t_i^a$, $t_{ij}^{ab}$, and $t_{XYZ}$) are found by projecting $e^{-T_{\rm SVD}} H 
e^{T_{\rm SVD}} |\phi_0\rangle=0$ onto a proper subset of excited determinants, and 
solving the resulting non-linear equations. The dimension of the compressed 
amplitudes tensor $t_{XYZ}$ is denoted by $N_{\mathrm{SVD}}$, see Table~\ref{tab:notation}, 
and it scales linearly with the system size. Note that this tensor is supersymmetric, i.e. 
invariant to any permutation of the indices $X$, $Y$, $Z$.

As a computationally convenient representation of the electron repulsion integrals we employ the 
density-fitting approximation~\cite{whitten73,baerends73,dunlap79,alsenoy88,vahtras93}
\begin{align}
\label{df}
 (pq|rs) \approx B_{pq}^Q\,B_{rs}^Q,\;\;\;\mbox{with}\;\;\;B_{pq}^Q = 
(pq|P)\,[\mathbf{V}^{-1/2}]_{PQ},
\end{align}
where $(pq|P)$ and $V_{PQ}=(P|Q)$ are the three-center and two-center electron repulsion 
integrals, respectively. Because the Coulomb metric is used in Eq. (\ref{df}) for determination 
of density-fitting coefficients, this formula is automatically ``robust'' in the sense that the 
error in the integrals is quadratic in the density errors~\cite{dunlap79}. The capital letters $P$, 
$Q$ are employed in the present work for the elements of the auxiliary basis set. The number of 
auxiliary basis set functions is denoted by the symbol $N_{\mathrm{aux}}$. By 
construction, $N_{\mathrm{aux}}$ scales linearly with the size of the system. In all calculations reported in this work, the error in relative energies caused by the density-fitting approximation was negligible in comparison with other uncertainties. This is consistent with other studies on this topic found in the literature~\cite{deprince13,lesiuk20b}.

\subsection{Non-iterative quadruples corrections}
\label{subsec:nonitt4}

The problem of economical inclusion of quadruple excitations effects in the CC theory was 
first considered by Kucharski, Bartlett and collaborators~\cite{kucharski86,kucharski89,kucharski89b,noga89,bartlett90,kucharski98,kucharski98b}. They introduced a non-iterative method, denoted 
CCSDT[Q] or simply [Q] in the present work, based on the standard M\o{}ller-Plesset perturbation 
theory where the Hartree-Fock determinant serves the role of the zeroth-order wavefunction. The quadruple excitation cluster operator $T_4$ is obtained from an approximate formula
\begin{align}
\label{t43a}
 \langle \mu_4 | \big[ F, T_4\big] + \big[ W, 
 T_3 \big] + \half \Big[\big[W, T_2\big],T_2\Big] \rangle = 0,
\end{align}
where $\mu_4$ stands for an appropriate string of quadruple excitation operators, i.e. $\mu_4=E_{ai}\,E_{bj}\,E_{ck}\,E_{dl}$, and hence $\langle \mu_4 |$ denotes projection onto the quadruply-excited configurations. A similar notation is used below also for lower-order excitations, e.g. $\mu_3=E_{ai}\,E_{bj}\,E_{ck}$. As the Fock operator is diagonal in the canonical orbital basis, Eq. (\ref{t43a}) can be explicitly solved to get the quadruply excited amplitudes in a closed-form
\begin{align}
 \label{t43b}
 t_{ijkl}^{abcd} = (\epsilon_{ijkl}^{abcd})^{-1} \langle \mu_4 | \big[ W, 
 T_3 \big] + \half \Big[\big[W, T_2\big],T_2\Big] \rangle,
\end{align}
where $\epsilon_{ijkl}^{abcd} 
=\epsilon_i+\epsilon_j+\epsilon_k+\epsilon_l-\epsilon_a-\epsilon_b-\epsilon_c-\epsilon_d$ is the 
four-particle energy denominator. The CCSDT[Q] correction to the energy (abbreviated as $E_{\rm 
[Q]}$) originating from the missing quadruple excitations then reads
\begin{align}
\label{e5q}
 E_{\rm [Q]} \equiv E_{\rm Q}^{[5]} = \langle T_2 | \big[ W, T_4\big] \rangle,
\end{align}
where the superscript $[5]$ indicates that the term enters in the fifth order of the M\o{}ller-Plesset 
perturbation theory. An alternative method to account for the quadruple excitations was presented 
by Bomble \emph{et al.}~\cite{bomble05} who employed L\"{o}wdin's partitioning of the coupled-cluster EOM 
Hamiltonian~\cite{stanton93}. This method is nowadays most commonly referred to as CCSDT(Q). The main difference 
between this approach and the pioneering developments of Kucharski and Bartlett is that the CCSDT 
wavefunction, rather than the Hartree-Fock determinant, is employed as zeroth-order wavefunction. 
The resulting energy correction, denoted by $E_{\rm (Q)}$, uses the same formula (\ref{t43b}) for 
the quadruply-excited amplitudes, but is given by sum of two terms
\begin{align}
\label{eqtot}
 E_{\rm (Q)} = E_{\rm Q}^{[5]} + E_{\rm Q}^{[6]}.
\end{align}
The former term is the same as in the CCSDT[Q] method, Eq. (\ref{e5q}), while the latter reads
\begin{align}
\label{e6q}
 E_{\rm Q}^{[6]} = \langle T_3 | \big[ W, T_4\big] \rangle.
\end{align}
As suggested by the notation, the term $E_{\rm Q}^{[6]}$ is of the sixth order in the usual 
perturbation theory and hence it was neglected in the CCSDT[Q] method. However, it has been shown~\cite{bomble05} that the importance of the $E_{\rm Q}^{[6]}$ contribution is much larger than its formal 
order would suggest. Only in the basis sets of double-zeta quality the term $E_{\rm Q}^{[5]}$ is 
dominating and the contribution from $E_{\rm Q}^{[6]}$ is typically smaller by an order of 
magnitude. This changes when the size of the basis set is increased to triple-zeta or larger. The 
terms $E_{\rm Q}^{[5]}$ and $E_{\rm Q}^{[6]}$ are then of a similar magnitude, with the latter 
even becoming dominant in some cases. As an immediate consequence, relative accuracy of the 
CCSDT[Q] method (in comparison to CCSDTQ) deteriorates with increasing basis set size. The accuracy 
level of the (Q) correction, on the other hand, was found to be remarkably consistent at least up to 
quintuple-zeta basis sets~\cite{bomble05}. In the present work we concentrate primarily on the implementation of the CCSDT(Q) method as a way of incorporating the effects of quadruple excitations into the rank-reduced CC formalism. To further justify this choice, below we provide a short survey of other available methods. We concentrate on closed-shell systems and hence open-shell generalizations are not discussed.

In the factorizable [Q] method~\cite{kucharski98}, denoted accordingly by [Q$_{\rm f}$], one employs the factorization 
theorem~\cite{frantz60} to get rid of the four-particle denominator in evaluation of the $E_{\rm 
Q}^{[5]}$ term. While this factorization is only approximate in the case of CCSDT amplitudes entering Eq. 
(\ref{e5q}), the quality of this approximation is usually excellent. The main advantage of the 
[Q$_{\rm f}$] method is the reduced scaling of the computational cost with the system size. While 
the exact computation of the [Q] correction scales as $N^9$, evaluation of the factorizable variant 
[Q$_{\rm f}$] can be accomplished with the $N^7$ cost. Unfortunately, the [Q$_{\rm f}$] method 
itself is an approximation to the [Q] correction, and hence it is bound to suffer from the same 
basis set dependency problems. To the best of our knowledge, analogous factorization cannot be 
accomplished for the $E_{\rm Q}^{[6]}$ term that involves projection onto the triply excited 
amplitudes.

The second family of non-iterative quadruples corrections retains different parametrization of the 
left- and right-hand-side coupled-cluster wavefunctions~\cite{kallay05}. The resulting CCSDT[Q]$_\Lambda$ and CCSDT(Q)$_\Lambda$ methods offer a noticeable improvement in terms of the accuracy in comparison to 
their conventional counterparts described above. However, this comes at a cost of evaluating the 
so-called coupled-cluster Lagrangian which is not available at present for the rank-reduced CCSDT 
method and requires a separate study. Next, we discuss the recently introduced CCSDT(Q-$n$) family of methods derived from Lagrangian-based perturbation theory, treating CCSDT as the 
zeroth-order wavefunction~\cite{eriksen14a,eriksen14b}. This framework is free from size-inconsistency problems encountered in 
the preceding EOM-like approaches and has been shown to converge rapidly to the exact 
CCSDTQ limit. While CCSDT(Q-$2$) is not competitive with the CCSDT(Q) theory, the improved
CCSDT(Q-$3$) variant offers an excellent accuracy level~\cite{eriksen15}. Unfortunately, the computational cost of 
CCSDT(Q-$3$) method is comparable to a single CCSDTQ iteration ($N^{10}$ scaling) and hence it is 
beyond the scope of the present work. Last but not least, the renormalized and completely renormalized approaches developed by Piecuch and collaborators~\cite{kowalski00,piecuch02,piecuch04,piecuch05} are derived using the so-called CC method of moments. They drastically improve the accuracy for systems with significant multireference character, but for systems dominated by a single reference determinant the results are similar.

\subsection{Quadratic (Q) functional: exact formulation}
\label{subsec:quadratic1}

In the rank-reduced context, the formulation of the (Q) correction based on Eqs. (\ref{e5q}), (\ref{eqtot}), and (\ref{e6q}) has a significant disadvantage. It stems from the fact that these equations were derived assuming that the $T_1$, $T_2$, and $T_3$ amplitudes come from the exact CCSDT theory, and $T_4$ amplitudes are obtained by solving Eq. (\ref{t43a}) without further approximations. In the rank-reduced formalism these assumptions do not hold; for example, the $T_3$ amplitudes are subject to the Tucker-$3$ compression, see Eq. (\ref{t3}). Unfortunately, if approximate cluster amplitudes are used to evaluate Eqs. (\ref{e5q}) and (\ref{e6q}), the error in the (Q) correction is roughly proportional to the error of the amplitudes. In other words, there exists an approximate linear relationship connecting the average error in the amplitudes and the error in the (Q) correction.
To avoid this problem and to guarantee that the latter error vanishes more rapidly as the accuracy of the amplitudes is improved, we propose a different functional for evaluation of the (Q) correction. The general form of the new functional, denoted $\mathcal{L}_{\rm (Q)}$ further in the text, reads
\begin{align}
\label{mainlq}
\begin{split}
 \mathcal{L}_{\rm (Q)} &= \langle T_2 | \big[ W, T_4\big] \rangle + \langle T_3 | \big[ W, T_4\big] \rangle
 + \langle L_3 | \,e^{-T} H e^{T} \rangle \\
 &+ \langle L_4 | \big[ F, T_4\big] + \big[ W, T_3 \big] + \half \Big[\big[W, T_2\big],T_2\Big] \rangle.
\end{split}
\end{align}
where $L_3$ and $L_4$ are two new auxiliary operators which assume the standard form
\begin{align}
 L_3 = \frac{1}{6}\, l_{ijk}^{abc}\,\mu_3,\;\;\;\mbox{and}\;\;\;
 L_4 = \frac{1}{24}\, l_{ijkl}^{abcd}\,\mu_4,
\end{align}
and the new amplitudes $l_{ijk}^{abc}$ and $l_{ijkl}^{abcd}$ are yet to be determined. 
The proposed functional has to fulfill two main theoretical requirements in order to be useful in the rank-reduced context:
\begin{itemize}
 \item if the exact CCSDT amplitudes together with $T_4$ amplitudes calculated from Eq. (\ref{t43a}) are used, the new functional gives \emph{strictly} identical results as the original formulation based on Eqs. (\ref{e5q}) and (\ref{e6q});
 \item the error of the (Q) correction evaluated using the new functional is quadratic in the error of the $T_3$/$L_3$ and $T_4$/$L_4$ amplitudes.
\end{itemize}
The motivation behind the first requirement is to enforce that in the limit of the complete triple excitation subspace 
in Eq. (\ref{t3}), i.e. when the SVD-CCSDT method is equivalent to the conventional CCSDT, the exact (Q) correction is 
recovered. Regarding the second requirement, the goal is to reduce the impact of the approximations adopted in the 
treatment of $T_3$ and $T_4$ amplitudes on the accuracy of the (Q) correction. However, one might ask why the quadratic 
error property is enforced only with respect to the $T_3$ and $T_4$ amplitudes, disregarding the $T_2$ amplitudes that 
enter Eq. (\ref{e5q}) directly and Eq. (\ref{e6q}) indirectly \emph{via} the $T_4$ operator. The justification is purely 
pragmatic and is based on a numerical observation that the $T_2$ amplitudes obtained from SVD-CCSDT method are 
sufficiently accurate for the purposes of evaluating Eqs. (\ref{e5q}) and (\ref{e6q}). In fact, we verified that even 
the use of CCSD $T_2$ amplitudes results in acceptable errors. This finding is not entirely surprising as similar 
arguments are used in the derivation of the aforementioned factorizable approximation to the $E_{\rm Q}^{[5]}$ term. All 
in all, while the second requirement given above can be strengthened to include the $T_2$ amplitudes as well, we found 
no practical reason to justify such choice, taking into account the increased complexity of the resulting formalism.

It is straightforward to verify that for any $L_3$ and $L_4$ operators, the $\mathcal{L}_{\rm (Q)}$ functional automatically fulfills the first requirement given in the previous paragraph. In fact, when the exact CCSDT amplitudes are inserted into the above formula, the third term vanishes identically as a consequence of the CCSDT stationary condition for the triple excitation amplitudes, $\langle \mu_3|\,e^{-T} H e^{T}\rangle=0$. Similarly, if the $T_4$ amplitudes are determined from Eq. (\ref{t43a}) without approximations, the fourth term included in $\mathcal{L}_{\rm (Q)}$ also vanishes, as it is a projection of Eq. (\ref{t43a}) onto some set of $L_4$ amplitudes.

In order to satisfy the second requirement discussed above, we demand that the $\mathcal{L}_{\rm (Q)}$ functional is stationary with respect to variations in the $T_3$, $T_4$, $L_3$, and $L_4$ amplitudes. In other words, we impose a condition that the first derivative of Eq. (\ref{mainlq}) with respect to each of these amplitudes separately is zero. Differentiation with respect to the $L_3$ and $L_4$ amplitudes returns back the stationary conditions and Eq. (\ref{t43a}), respectively, and hence this brings no new information into the formalism. By differentiating $\mathcal{L}_{\rm (Q)}$ with respect to the $T_3$ and $T_4$ amplitudes, respectively, and setting the resulting equations to zero one obtains
\begin{align}
\label{eql3}
 \langle \mu_3|\big[ W, T_4\big] \rangle +
 \langle L_3 | \Big[e^{-T}H\,e^{T},\mu_3\Big]  \rangle +
 \langle L_4 | \big[ W, \mu_3 \big]\rangle = 0,
\end{align}
and
\begin{align}
\label{eql4}
 \langle T_2 | \big[ W, \mu_4\big] \rangle +
 \langle T_3 | \big[ W, \mu_4\big] \rangle +
 \langle L_4 | \big[ F, \mu_4\big] \rangle = 0,
\end{align}
respectively. The latter equation can be directly solved to obtain the $L_4$ amplitudes. Upon inserting the results into Eq. (\ref{eql3}), it becomes a system of linear equations with the $L_3$ amplitudes being the only unknowns. Therefore, Eqs. (\ref{eql3}) and (\ref{eql4}) completely determine the auxiliary $L_3$ and $L_4$ operators, and hence enable calculation of the modified (Q) functional, Eq. (\ref{mainlq}).

With Eqs. (\ref{eql3}) and (\ref{eql4}) at hand, it remains to show that the quantity $\mathcal{L}_{\rm (Q)}$ indeed 
fulfills the second condition discussed above. The simplest way to achieve this is to employ the chain rule of 
differentiation and exploit the stationary conditions (\ref{mainlq}). However, in Supporting Information we provide a 
more detailed derivation that has the advantage of providing a rigorous error estimation, namely
\begin{align}
\label{l2}
 \begin{split}
 \delta\mathcal{L}_{\rm (Q)} &= 
 \langle \delta T_3 | \big[ W, \delta T_4\big] \rangle
 + \langle \delta L_3 | \,\Big[e^{-T^{\mathrm{ex}}} H e^{T^{\mathrm{ex}}}, \delta T_3\Big]\rangle
 + \langle \delta L_4 | \big[ F, \delta T_4\big] + \big[ W, \delta T_3 \big]\rangle,
\end{split}
\end{align}
where $T^{\mathrm{ex}}$ denotes the exact CCSDT amplitudes, while $\delta T_3$ is an error in the $T_3$ operator 
(analogous notation is used for the remaining quantities). It is straightforward to verify that each term of the above 
formula is quadratic in the combined powers of $\delta T_3$, $\delta L_3$, $\delta T_4$, and $\delta L_4$. This proves 
that the proposed functional $\mathcal{L}_{\rm (Q)}$ satisfies the second requirement introduced at the beginning of 
this section.

\subsection{Quadratic (Q) functional: approximations}
\label{subsec:quadratic2}

In order to make calculations based on the quadratic (Q) functional feasible, approximations need to be introduced to 
the exact formalism presented in the previous section. However, we stress that in order to retain the desirable 
properties of the $\mathcal{L}_{\rm (Q)}$ functional, no approximations are made to its formal definition given by Eq. 
(\ref{mainlq}). Instead, we adopt several simplifications to the equations that determine the $T_3$, $T_4$, $L_4$, and 
$L_3$ operators, as described in detail in this section. Due to the quadratic nature of the $\mathcal{L}_{\rm (Q)}$ 
functional, these approximations are expected to have a small impact on the accuracy of the (Q) correction.

Starting with the $T_3$ operator, it is given by the approximate form (\ref{t3}), inherited after the SVD-CCSDT theory, with the amplitudes compressed using the Tucker-$3$ format. We adopt no further approximations to this quantity.

Moving to the $T_4$ operator, the handling and storage of the full-rank quadruply-excited amplitudes given by Eq. (\ref{t43b}) constitutes the major bottleneck of the exact formalism. To overcome this obstacle we approximately cast the quadruply-excited amplitudes (\ref{t43b}) in a rank-reduced form employing the Tucker format
\begin{align}
 \label{tucker4}
 t_{ijkl}^{abcd} \approx t_{ABCD} \,V^A_{ai}\,V^B_{bj}\,V^C_{ck}\,V^D_{dl},
\end{align}
which is fully analogous to the rank-reduced form of the triply-excited amplitudes, cf. Eq. (\ref{t3}). As the expansion 
basis vectors $V^A_{ai}$ are distinct from their counterparts used in Eq. (\ref{t3}), we employ the capital letters 
$A,B,C,\ldots$ to denote the quantities that relate to the quadruply-excited amplitudes. The dimension of the core 
tensor $t_{ABCD}$ in Eq. (\ref{tucker4}) is referred to as $N_{\rm qua}$. By analogy with the 
findings for the doubly- and triply-excited amplitudes represented in the Tucker-$n$ format, we assume that $N_{\rm 
qua}$ scales linearly with the system size. A numerical demonstration of this condition is presented in 
Sec.~\ref{subsec:scaling}. The conversion of the full-rank quadruply-excited amplitudes $t_{ijkl}^{abcd}$ to 
the compressed form (\ref{tucker4}) is non-trivial. To accomplish this task we propose a novel algorithm based on 
higher-order orthogonal iteration (HOOI). Details of this procedure are described in the next section, along with 
the analysis of the computational costs and scaling with the system size.

Next, we consider the auxiliary operator $L_4$ which is defined by Eq.~(\ref{eql4}). To facilitate efficient evaluation 
of this quantity, we adopt two levels of approximatiofns. First, in Eq.~(\ref{eql4}) we neglect the term that involves 
the triply excited amplitudes, namely $\langle T_3 | \big[ W, \mu_4\big] \rangle$. 
The justification of this approximation is rooted in the standard M\o{}ller-Plesset perturbation theory, where the $T_2$ operator enters in the first-order perturbed wavefunction, while the $T_3$ operator appears 
in the second order. By the same token, we expect the contribution of the $\langle T_2 | \big[ W, \mu_4\big] \rangle$ to be dominating, while the neglected term constitutes a relatively minor correction. The neglect of the term $\langle T_3 | \big[ W, \mu_4\big] \rangle$ leads to the modified expression
\begin{align}
\label{eql4_app}
 \langle T_2 | \big[ W, \mu_4\big] \rangle + \langle L_4 | \big[ F, \mu_4\big] \rangle = 0,
\end{align}
which can easily be solved directly by exploiting the diagonal nature of the canonical Fock operator, giving
\begin{align}
 \label{eql4_app_solved}
 l_{ijkl}^{abcd} = (\epsilon_{ijkl}^{abcd})^{-1} \langle \mu_4 |\,W\,T_2\rangle.
\end{align}
In Supporting Information we provide an explicit formula for this quantity expressed through the basic CC amplitudes and two-electron integrals. The second level of approximation adopted for $L_4$ is the same as for the $T_4$, i.e. rank-reduction to the Tucker-$4$ format
\begin{align}
 \label{lucker4}
 l_{ijkl}^{abcd} \approx l_{A'B'C'D'} \,V^{A'}_{ai}\,V^{B'}_{bj}\,V^{C'}_{ck}\,V^{D'}_{dl},
\end{align}
where the primes have been added to underline that the expansion basis is different from that of Eq. (\ref{tucker4}). Similarly as for the $T_4$ operator, in the next section we provide technical details of the HOOI procedure used to determine the rank-reduced $L_4$ amplitudes.

Finally, let us consider the $L_3$ operator for which the approximation scheme is somewhat more involved and consists of two steps. In the first step, we neglect the fluctuation potential $W$ in the similarity-transformed Hamiltonian present in Eq. (\ref{eql3}) and set $e^{-T}H\,e^{T}\approx e^{-T}F\,e^{T}$. This leads to the modified formula
\begin{align}
\label{eql3_app1}
 \langle \mu_3|\big[ W, T_4\big] \rangle +
 \langle L_3 | \Big[F,\mu_3\Big]  \rangle +
 \langle L_4 | \big[ W, \mu_3 \big]\rangle = 0,
\end{align}
where we have additionally exploited the fact that $\big[e^{-T}F\,e^{T},\mu_3\big]=\big[F,\mu_3\big]$ which is straightforward to prove using the BCH expansion. By exploiting the properties of the Fock operator, explicit solution of this equation is written as
\begin{align}
 \label{eql3_solved}
 l_{ijk}^{abc} = (\epsilon_{ijk}^{abc})^{-1} \langle \mu_3 |\big[ W, T_4+L_4\big]\rangle,
\end{align}
where the equality $\langle L_4 | \big[ W, \mu_3 \big]\rangle=\langle \mu_3|\big[ W, L_4\big] \rangle$ has been used for convenience sake.

To introduce the second layer of approximation, we note that the third term in Eq. (\ref{eql3_app1}) is dominating in comparison with the first term. This is again justified by perturbation theory arguments; by comparing Eqs. (\ref{eql4_app}) and (\ref{t43a}) we see that $L_4$ is a second-order quantity, while $T_4$ is a third-order quantity. Therefore, it is tempting to neglect the $\langle \mu_3|\big[ W, T_4\big] \rangle$ term altogether in Eq. (\ref{eql3_app1}), in a similar spirit as in the previous paragraph where approximations of the $L_4$ operator were discussed. We considered this approach in the preliminary stage of the implementation and verified that it indeed delivers a decent accuracy level. However, there exists an alternative approach to approximating the $L_3$ operator which is based on the Tucker format
\begin{align}
 \label{l3_hooi}
 l_{ijk}^{abc} \approx l_{X'Y'Z'} \,U^{X'}_{ai}\,U^{Y'}_{bj} \,U^{Z'}_{ck},
\end{align}
where the primes indicate that the quantities $U^{X'}_{ai}$ are distinct from the expansion basis used for the triply-excited amplitudes in Eq. (\ref{t3}). Similarly as for other quantities, HOOI is used to bring $l_{ijk}^{abc}$ into the decomposed form (\ref{l3_hooi}). However, as a cost-saving measure, we introduce a simplification: the expansion basis $U^{X'}_{ai}$ is found by decomposing an approximate form of the $l_{ijk}^{abc}$, namely
\begin{align}
 \label{eql3_solved_app}
 l_{ijk}^{abc} \approx (\epsilon_{ijk}^{abc})^{-1} \langle \mu_3 |\big[ W, L_4\big]\rangle,
\end{align}
where the term including the $T_4$ operator has been neglected, cf. Eq. (\ref{eql3_solved}). Once the quantities $U^{X'}_{ai}$ are found, the core tensor $l_{X'Y'Z'}$ is found by projection:
\begin{align}
 l_{X'Y'Z'} = (\epsilon_{ijk}^{abc})^{-1} \langle \mu_3 |\big[ W, T_4+L_4\big]\rangle\,U^{X'}_{ai}\,U^{Y'}_{bj} \,U^{Z'}_{ck}.
\end{align}
Note that in this step the full form of the $L_3$ operator is used, without any approximations in comparison with Eq. 
(\ref{eql3_solved}). We found that this hybrid approach reduces the computational cost of decomposing the amplitudes 
$l_{ijk}^{abc}$ considerably without affecting the accuracy of the decomposition (\ref{l3_hooi}).
This can be explained by noting that the term involving the $T_4$ is numerically minor. Therefore, the basis found using 
the decomposition of the approximate formula, Eq. (\ref{l3_hooi}), is able to accommodate both terms accurately, despite 
the $T_4$ is absent in the optimization procedure.

To study the impact of the proposed approximations, we carried out calculations for small molecular systems from Sec.~\ref{subsec:param}. In Supporting Information we provide detailed results for the most representative case using the cc-pVDZ, cc-pVTZ, and cc-pVQZ basis sets. These data show that the neglected terms are numerically small and some additional approximations proposed above have a small impact on the accuracy of the proposed formalism.

Finally, let us point out that, in general, the dimensions of the quadruple excitation subspace ($\nqua$) used for compression of the $t_{ijkl}^{abcd}$ and $l_{ijkl}^{abcd}$ amplitudes, i.e. in Eqs. (\ref{tucker4}) and (\ref{lucker4}), can be different. Similarly, different sizes of the triple excitation subspace ($\nsvd$) may be used for $T_3$ and $L_3$ operators. However, in a set of preliminary calculations we found that near-optimal results are attained when the same size of the quadruple excitation subspace is used for $T_4$ and $L_4$, and the same size of triple excitation subspace for $T_3$ and $L_3$. Therefore, we use a single parameter $\nqua$ to denote the length of the expansion in both Eq.~(\ref{tucker4})~and~(\ref{lucker4}), and similarly $\nsvd$ for both Eq.~(\ref{l3_hooi}) and (\ref{t3}). Gains in terms of accuracy achieved by lifting these restrictions do not justify the corresponding increase of the technical complexity of the formalism.

\subsection{Compression of the excitation amplitudes}
\label{subsec:t4hooi}

As mentioned in the previous section, the decomposition of the triply- and quadruply-excited amplitudes, required to 
evaluate the approximate form of the quadratic functional detailed in Sec.~\ref{subsec:quadratic2}, is achieved using 
the higher-order orthogonal iteration (HOOI) procedure. In this section we provide details of this procedure, both theoretical 
and technical. We concentrate primarily on the decomposition of the $t_{ijkl}^{abcd}$ tensor as this is the most 
problematic quantity. However, extension of this procedure to the amplitudes present in the $L_3$ and $L_4$ 
operators is briefly discussed at the end of the present section, and further technical details are presented in Supporting Information.

In the HOOI procedure the decomposition of the amplitudes is achieved by minimization of the 
following cost function
\begin{align}
 \label{lsquares}
 \sum_{ijkl} \sum_{abcd} \Big[
  t_{ijkl}^{abcd} - t_{ABCD} \,V^A_{ai}\,V^B_{bj}\,V^C_{ck}\,V^D_{dl}
  \Big]^2,
\end{align}
subject to the condition that the $V^A_{ai}$ vectors are column orthonormal, that is
\begin{align}
\label{vortho}
 V^A_{ai}\,V^B_{ai} = \delta_{AB}.
\end{align}
The constraint (\ref{vortho}) is imposed without loss of generality, because any linear 
transformation amongst the $V^A_{ai}$ vectors can be counteracted by changing the values of the 
core tensor, leaving the cost function unaffected. For a fixed expansion length ($N_{\rm 
qua}$) the least-squares problem (\ref{lsquares}) is solved by HOOI which in the present case proceeds as follows. Assuming that an 
approximate solution $V^A_{ai}$ of the minimization problem (\ref{lsquares}) is known, one 
forms an intermediate quantity
\begin{align}
\label{halfp}
 t_{ai,BCD} = t_{ijkl}^{abcd}\,V^B_{bj}\,V^C_{ck}\,V^D_{dl},
\end{align}
which is a partial projection of the full-rank tensor onto the current subspace. Next, the truncated
singular value decomposition (SVD) of the $t_{ai,BCD}$ tensor is computed. The left-singular 
vectors corresponding to $N_{\rm qua}$ the largest singular values form the updated expansion 
vectors, $V^A_{ai}$. Note that by the virtues of the SVD procedure the new vectors 
automatically obey the orthonormality constraint (\ref{vortho}). This basic iteration process is 
repeated until the convergence criterion is met; a convenient choice of the stopping criteria is 
discussed further in the text. Note that during the HOOI procedure the core tensor $t_{ABCD}$ does 
not have to be formed explicitly. Nonetheless, it is obtained straightforwardly as
\begin{align}
\label{corefin}
 t_{ABCD} = t_{ai,BCD}\,V^A_{ai},
\end{align}
owning to the orthonormality of the expansion vectors, $V^A_{ai}$.

The basic HOOI procedure described above constitutes a serviceable method. Unfortunately, due to 
large dimensions of the $t_{ai,BCD}$ matrix, namely $OV\times N_{\rm qua}^3$, the SVD step of this 
algorithm is too expensive for large-scale applications. Therefore, we introduce a modification of 
the HOOI procedure where instead of the $t_{ai,BCD}$ matrix the following quantity is computed
\begin{align}
\label{mmatrix}
 M_{ai,bj} = t_{ai,BCD} \, t_{bj,BCD}.
\end{align}
Let us recall that for any rectangular matrix $\mathbf{M}$, its left singular-vectors coincide with 
the eigenvectors of the normal matrix $\mathbf{M}\mathbf{M}^{\mathrm{T}}$. Therefore, the
updated expansion vectors $V^A_{ai}$ can equivalently be obtained by diagonalizing the 
$M_{ai,bj}$ matrix and retaining the eigenvectors corresponding to the largest eigenvalues (which 
are non-negative by construction). The dimensions of the $M_{ai,bj}$ matrix are $OV\times OV$ and 
hence it admits eigendecomposition in $N^6$ time, a significant improvement over the SVD of the 
$t_{ai,BCD}$ matrix. As a by-product, the modification of the HOOI procedure described above solves 
the memory bottleneck related to the storage of the complete $t_{ai,BCD}$ matrix ($N^5$ memory 
chunk). In our implementation, the $t_{ai,BCD}$ matrix is calculated in batches with one of the 
$B,C,D$ indices fixed. The batches are then immediately used to compute the contribution to 
$M_{ai,bj}$ without accumulation of the full $t_{ai,BCD}$ matrix. The storage requirements are 
reduced in this way to the level of $OVN_{\rm qua}^2\propto N^4$. 

It remains to discuss two technical aspects of the HOOI algorithm, i.e., the choice of the stopping 
criteria and the starting values. As discussed at length in Ref.~\onlinecite{lesiuk22} an appropriate stopping 
condition is obtained by monitoring the norm of the core tensor
\begin{align}
\label{tnorm}
 ||t||^2 = \sum_{ABCD} t_{ABCD}^2 = \sum_A M_{ai,bj}\,V^A_{ai}\,V^A_{bj},
\end{align}
where the second equality follows from the orthonormality of the $V^A_{ai}$ vectors. The HOOI 
procedure is ended when the relative difference in $||t||$ between two consecutive iterations falls 
below a predefined threshold, $\epsilon$. The threshold value of $\epsilon=10^{-6}$ is sufficient 
in most applications and has been adopted in the present work. The cost of computing $||t||$ is 
negligible in comparison with other parts of the algorithm. The reason why this straightforward 
procedure performs well in practice is a consequence of the fact that the HOOI algorithm can be 
reformulated as a maximization of the norm of the core tensor instead of minimization of Eq. 
(\ref{lsquares}), see Refs.~\onlinecite{delath00b,sun22}.

The problem of the starting values is solved simply by setting $V^A_{ai}$ as equal to 
$U^X_{ai}$ that correspond to the largest absolute diagonal values of $t_{XYZ}$, see Eq. 
(\ref{t3}). We found this procedure to be entirely satisfactory in practice and 
convergence of the HOOI procedure is achieved typically within only $5-10$ iterations to accuracy 
of $\epsilon=10^{-6}$.
The sole exception from this rule occurs in calculations with extremely small $N_{\rm qua}$, but it 
is questionable whether they are of any practical importance. The major steps of the HOOI algorithm 
are summarized in Algorithm~\ref{alghooi}.

\begin{algorithm}[ht!]
\SetAlgoLined
 generate starting values for $V_{ai}^A$\;
 set $||t||_0=0$ and conv. threshold $\varepsilon$\;
 \For{$n=1$ \KwTo $\rm{maxit}$}{
  set $M_{ai,bj}=0$\;
  \For{$D=1$ \KwTo $N_{\rm qua}$}{
   compute $t_{ai,BCD}$ with fixed $D$ index\;
   $t_{ai,BCD}\,t_{bj,BCD} \rightarrow M_{ai,bj}$, accumulate\;
  }
  compute $||t||_n$, see Eq. (\ref{tnorm})\;
  if $||t||_n - ||t||_{n-1} < \varepsilon$ then exit\;
  set $||t||_{n-1} = ||t||_n$\;
  diagonalize $M_{ai,bj}$\;
  find $N_{\rm qua}$ eigenvectors with the largest \\
  eigenvalues and place them in updated $V_{ai}^A$\;
 }
 \caption{\label{alghooi} Pseudo-code for the higher-order orthogonal iteration (HOOI) algorithm
          applied to the quadruply-excited amplitudes.}
\end{algorithm}

According to the above discussion, the critical part of the HOOI algorithm is the calculation of 
partly projected quadruply-excited amplitudes, Eq. (\ref{halfp}). Without invoking any 
approximations, the computational cost of this step is proportional to $N^9$, even under the 
assumption that $N_{\rm qua}$ scales linearly with the system size. This offers no 
practical advantages over the available conventional algorithms for calculation of the quadruples 
corrections. The main reason behind this steep scaling is the presence of the four-particle energy 
denominator in Eq. (\ref{t43b}) which has to be eliminated in order to enable any scaling reduction. 
To this end, we employ the discrete Laplace transformation (LT) technique
\begin{align}
\label{ltd4}
 (\epsilon_{ijkl}^{abcd})^{-1} = \sum_g^{N_g} w_g\,e^{-t_g \left( \epsilon_i^a + \epsilon_j^b + 
 \epsilon_k^c + \epsilon_l^d \right)},
\end{align}
where $t_g$ and $w_g$ are the quadrature nodes and weights, respectively, $N_g$ is the size of the 
quadrature, and $\epsilon_i^a = \epsilon_i - \epsilon_a$. For the two-particle energy denominator 
this method was first proposed by Alml\"{o}f~\cite{almlof91} in the context of the MP2 theory, but 
since then it has been successfully used in combination with other electronic structure methods~\cite{haser92,ayala99,lambert05,nakajima06,jung04,kats08}. In 
this work we employ the min-max quadrature proposed by Takatsuka and 
collaborators~\cite{takatsuka08,braess05,paris16} for the choice of $t_g$ and $w_g$. The number of 
quadrature points in Eq. (\ref{ltd4}) is independent of the system size, that is $N_g\propto N^0$. 
To the best of our knowledge, this is the first application of the LT technique to the four-particle 
energy denominator in the coupled-cluster theory.

The quadruply-excited amplitudes defined by Eq. (\ref{t43b}) are given by the following explicit 
expression
\begin{align}
 t_{ijkl}^{abcd} = (\epsilon_{ijkl}^{abcd})^{-1} \Gamma_{ijkl}^{abcd},
\end{align}
where
\begin{align}
\label{biggamma}
\begin{split}
 \Gamma_{ijkl}^{abcd} &= \half P_{ijkl}^{abcd} \bigg[ 
   (ai|be)\,t_{jkl}^{ecd}
 - (ai|mj)\,t_{mkl}^{bcd}
 + (nj|mi)\,t_{mk}^{ac}\,t_{nl}^{bd} \\
 &-2\,(ai|me)\,t_{kj}^{eb}\,t_{ml}^{cd}
  -2\,(be|mi)\,t_{kj}^{ce}\,t_{ml}^{ad}
  +(cf|ae)\,t_{ij}^{eb}\,t_{kl}^{fd}
  \bigg].
\end{split}
\end{align}
The permutation operator $P_{ijkl}^{abcd}$ in the above formula reads
\begin{align}
 P_{ijkl}^{abcd} = \left( 1 + P_{ai,bj} \right) \left( 1 + P_{ai,ck} + P_{bj,ck}\right)
 \left( 1 + P_{ai,dl} + P_{bj,dl} + P_{ck,dl}\right),
\end{align}
and $P_{ai,bj}$ denotes the basic transposition operator that exchanges pairs of indices 
$i\leftrightarrow j$ and $a\leftrightarrow b$ simultaneously. By employing the LT technique, we 
rewrite Eq. (\ref{halfp}) in the form
\begin{align}
\label{proj2}
 t_{ai,BCD} = \sum_g^{N_g} w_g\,e^{-t_g \epsilon_i^a}\,
 \Gamma_{ijkl}^{abcd}\,\hat{V}^{Bg}_{bj}\,\hat{V}^{Cg}_{ck}\,\hat{V}^{Dg}_{dl} = 
 \sum_g^{N_g} w_g\,e^{-t_g \epsilon_i^a}\,\gamma_{ai,BCD}^g,
\end{align}
with $\hat{V}^{Bg}_{bj} = V^B_{bj}\,e^{-t_g \epsilon_j^b}$ and 
$\gamma_{ai,BCD}^g=\Gamma_{ijkl}^{abcd}\,\hat{V}^{Bg}_{bj}\,\hat{V}^{Cg}_{ck}\,\hat{V}^{Dg}_{dl}$. 
Within this formulation of the problem, the overall scaling of assembling the quantity $t_{ai,BCD}$ 
can be reduced to the level of $N^6$. To illustrate this, let us consider a term $(bj|ae)\,t_{ikl}^{ecd}$ obtained by permutation of indices from the first term in Eq. (\ref{biggamma}). In the conventional implementation, this term scales as $O^4\,V^5$. However, a contribution of this term to the quantity 
$\gamma_{ai,BCD}^g$ required in Eq. (\ref{proj2}) can be factorized as
\begin{align}
\begin{split}
 &\gamma_{ai,BCD}^g \longleftarrow
 (bj|ae)\,t_{ikl}^{ecd}\,\hat{V}^{Bg}_{bj}\,\hat{V}^{Cg}_{ck}\,\hat{V}^{Dg}_{dl} = \\
 &\Big(\big(B_{ae}^Q\,U_{ei}^X\big)\,\big(B_{bj}^Q\,\hat{V}^{Bg}_{bj}\big)\Big)\,\Big[\Big(t_{XYZ}\,
 \big(U_{ck}^Y\,\hat{V}^{Cg}_{ck}\big)\Big)\,\big(U_{dl}^Z\,\hat{V}^{Dg}_{dl}\big)\Big],
\end{split}
\end{align}
where we have exploited the density-fitting factorization of the electron repulsion integrals, see 
Eq. (\ref{df}), and Tucker factorization of the triply-excited amplitudes tensor given by Eq. 
(\ref{t3}). The parentheses included in the above formula indicate the order of operations and 
should be read starting from the innermost bracket. By following the optimal order of contractions 
one can show that the most costly step scales as $N_g\,OV\nsvd\nqua^3\propto N^6$. Similar 
factorizations are possible for the remaining terms appearing in Eq. (\ref{biggamma}) and in every 
case the scaling is proportional to $N^6$, albeit with different prefactors. However, the number of 
terms in Eq. (\ref{biggamma}) that have to be factorized is large -- 144 in total if all 
permutations resulting from the action of the $P_{ijkl}^{abcd}$ are taken into account. Therefore, 
explicit factorized formulas for the quantity $\gamma_{ai,BCD}^g$ are given in Supporting 
Information, along with a detailed discussion of possible simplifications. 

Let us point out that the cost of the complete HOOI procedure is still asymptotically proportional 
to $N^7$ due to the need to assemble the $M_{ai,bj}$ matrix, see Eq. (\ref{mmatrix}). However, this 
step can be formulated as a single \textsc{dgemm} matrix-matrix multiplication and hence possesses 
a relatively small prefactor. Therefore, the calculation of the quantity $\gamma_{ai,BCD}^g$, in 
spite of the $N^6$ scaling, constitutes the majority of the total cost of the HOOI procedure for 
systems that can currently be studied. 

The presentation given in this section has been focused on the $t_{ijkl}^{abcd}$ amplitudes. However, the Algorithm~\ref{alghooi} is straightforwardly adapted for an analogous decomposition of the remaining quantities, namely the $l_{ijk}^{abc}$ and $l_{ijkl}^{abcd}$ amplitudes that parametrize the $L_3$ and $L_4$ operators. This is particularly seamless in the latter case as the only change required is the modification of Eq. (\ref{halfp}) without affecting any other steps of Algorithm~\ref{alghooi}. Efficient evaluation of the modified expression is discussed in Supporting Information and it is shown that the scaling of this step is $N^5$, i.e. lower than in the case of the $t_{ijkl}^{abcd}$ amplitudes. Somewhat more advanced modifications of the HOOI procedure are required for the $l_{ijk}^{abc}$ amplitudes. First, instead of Eq. (\ref{halfp}) one calculates $l_{ai,Y'Z'}=l_{ijk}^{abc} \,U^{Y'}_{bj} \,U^{Z'}_{ck}$, and the matrix $M_{ai,bj}$ required in Algorithm~\ref{alghooi} is obtained as
$M_{ai,bj}=l_{ai,Y'Z'}\,l_{bj,Y'Z'}$, cf. Eq.~(\ref{mmatrix}). The remaining steps of the HOOI procedure are not affected. In Supporting Information we provide a factorized expression that enables to compute the intermediate quantity $l_{ai,Y'Z'}$ with $N^6$ cost, meaning that the decomposition of the $l_{ijk}^{abc}$ amplitudes scales as $N^6$ overall.

To sum up the theoretical section of the paper, we point out that computation of the $\mathcal{L}_{\rm (Q)}$ functional involves several steps, some of which have to be performed in a predefined order. For completeness, these steps are summarized in the Supporting Information, identifying the relevant key equations and presenting additional technical details. Importantly, by analyzing the working expressions we show that all components of the $\mathcal{L}_{\rm (Q)}$ functional can be evaluated with a cost proportional to $N^7$ or less. A numerical confirmation of this finding is provided in the next section.

\section{Numerical results and discussion}
\label{sec:numer}

\subsection{Computational details}
\label{subsec:kompot}

Unless explicitly stated otherwise, in all calculations reported in this work employ the correlation-consistent cc-pV$X$Z basis set from Ref.~\onlinecite{dunning89}. The corresponding density-fitting auxiliary basis sets cc-pV$X$Z-RIFIT were taken from the work of Weigend et al~\cite{weigend02b}. In some calculations, specified further in the text, a larger cc-pV5Z-RIFIT basis from Ref.~\onlinecite{hattig05} was employed to minimize the density-fitting error. Pure spherical 
representation ($5d$, $7f$, etc.) of all Gaussian basis sets is adopted throughout. Density-fitting approximation is used in all correlated calculations unless written otherwise. However, the Hartree-Fock equations are solved using the exact two-electron integrals and hence the canonical HF orbitals are exact within the given one-electron basis. The frozen-core approximation is invoked in all computations reported in this work, unless explicitly stated otherwise. The $1s^2$ core orbitals of the first-row atoms (Li--Ne) are not correlated.

The reference CCSDT(Q) calculations and calculation of zero-point vibrational energies and harmonic frequencies were performed with the \textsc{CFour} program package~\cite{cfour1,cfour2}. Some of the more demanding CCSD(T) calculations reported in this work, indicated below, were performed using \textsc{NWChem} program\cite{nwchem20}, version 6.8. The calculations performed using the \textsc{CFour} and \textsc{NWChem} programs do not use the density-fitting approximations. All theoretical methods described in this work were implemented in a computer program written specifically for this purpose which is available from the author upon request. The \textsc{TBlis} library~\cite{matthews18} is used in the code for performing efficient tensor operations. It is worth mentioning that \textsc{TBlis} natively supports shared-memory multiprocessing (using \text{OpenMP} application programming interface in our case) and hence most of the calculations reported in this work are performed in parallel, unless specified otherwise. Speed-ups by approximately a factor of $10$ were observed in large-scale calculations on $12$ (internode) threads. Beyond this point overheads related to, e.g. load balancing and synchronization, become significant and further increase of the number of computing threads leads to diminishing returns. A higher level of parallelization is possible using the \text{MPI} standard. This requires to divide the workload into independent task and, in the present context, it is natural to distribute the $B_{pq}^Q$ three-center integrals by splitting the index $Q$ among the computing nodes. However, this possibility has not been exploited in the present work. Lastly, the current implementation of the proposed theory does not utilize spatial symmetry of the molecules. Therefore, all calculations reported in this work are performed within C$_1$ symmetry group.

To avoid confusion, we briefly touch upon the naming conventions used in this section. The abbreviation SVD-CCSDT designates the iterative rank-reduced CC method introduced in Ref.~\onlinecite{lesiuk20}, and described briefly in Sec.~\ref{subsec:pre}, which is based on the Tucker compression of the triple excitation amplitudes, Eq. (\ref{t3}). The abbreviation SVD-CCSDT+ refers to a method introduced in Ref.~\onlinecite{lesiuk21}. It consists of adding a non-iterative correction that accounts for triple excitations outside the subspace used in SVD-CCSDT calculations. In this way, the error with respect to the exact CCSDT method is reduced, even if the subspace of triple excitations used in Eq. (\ref{t3}) is small. Note that both SVD-CCSDT and SVD-CCSDT+ methods become functionally equivalent to the exact CCSDT for a sufficiently large value of the $N_{\mathrm{SVD}}$ parameter that controls the expansion length in Eq.~(\ref{t3}).

\subsection{Scaling demonstration}
\label{subsec:scaling}

The efficiency of the proposed method hinges on the assumption that the parameter $N_{\rm qua}$, which determines the size of the quadruple excitation subspace in Eq. (\ref{tucker4}), scales linearly with the system size, $N$. In other words, to retain a constant relative accuracy in the correlation energy as the system size grows, it is sufficient to set $N_{\rm qua}$ proportional to $N$. This conjecture is non-trivial because in the limit of the complete quadruple excitation subspace $N_{\rm qua}$ scales quadratically with the system size (more precisely, it is equal to the number of occupied times the number of virtual orbitals in the system).

\begin{figure}[ht]
\includegraphics[scale=0.75]{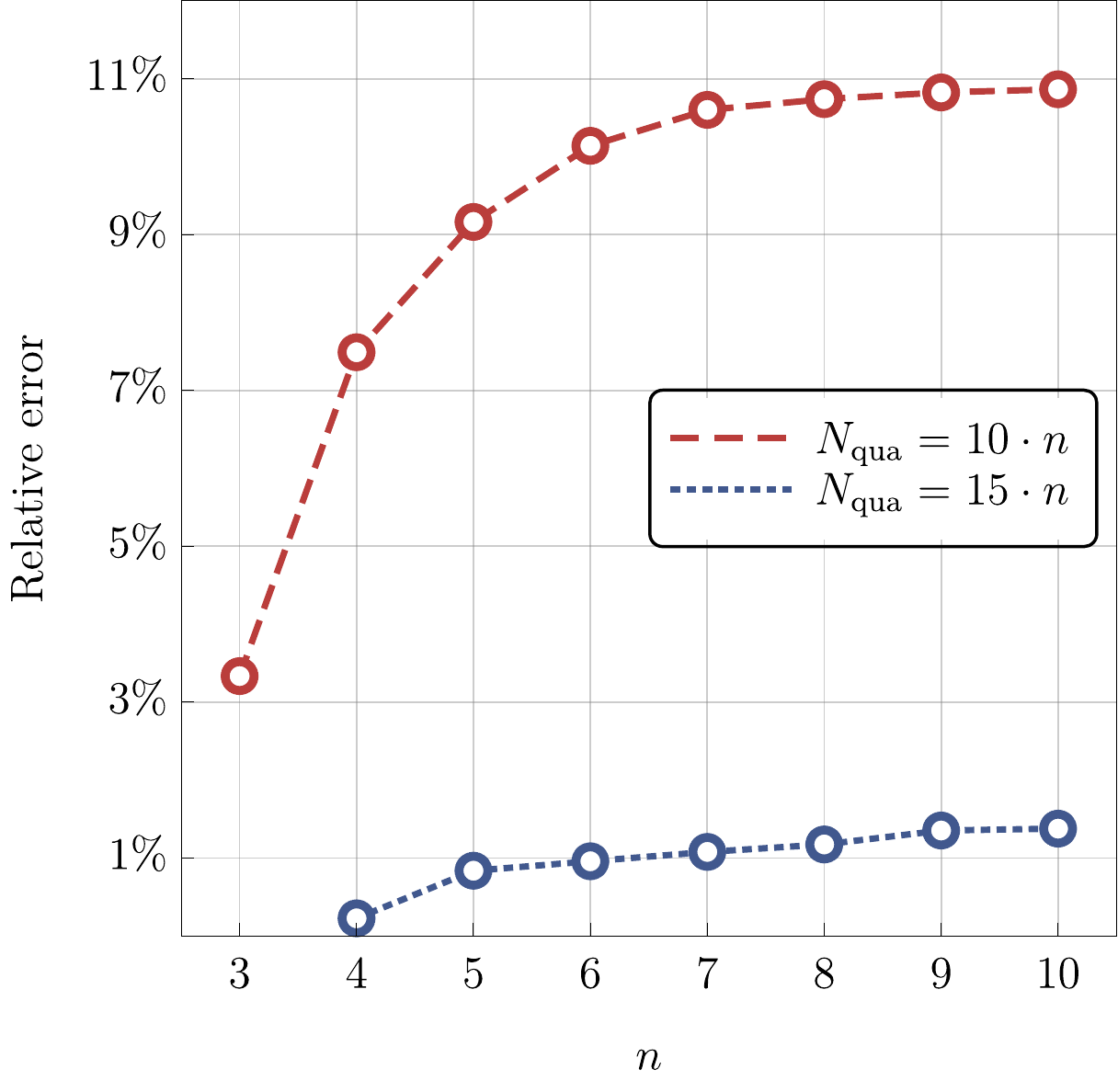}
\caption{\label{fig:error-scaling}
Percentage error in the (Q) correction obtained using the $\mathcal{L}_{\rm (Q)}$ functional 
(cc-pVDZ basis set) for the $(\mbox{H}_2)_n$ system, $n=3,\ldots,10$, with $N_{\rm qua}=10\cdot n$ and $N_{\rm 
qua}=15\cdot n$. The exact CCSDT(Q) results are used as a reference.
}
\end{figure}

In previous works, the property that dimension of the excitation subspace scales linearly with the system size
has been demonstrated numerically for lower-dimensional (two- and three-) analogs of Eq. (\ref{tucker4}) using realistic 
model systems such as linear alkanes or water clusters for which reference CCSD/CCSDT results are available~\cite{lesiuk20,parrish19,lesiuk22}. 
Unfortunately, evaluation of the (Q) correction for systems comprising more than $4-5$ non-hydrogen atoms is 
computationally costly, and hence reference results are not available for these model systems of sufficient size to 
reach definite conclusions about the behavior of $N_{\rm qua}$. As a compromise, we adopt linear hydrogen chains as a 
model systems for the purposes of the scaling demonstration. The chain is composed of H$_2$ molecules (bond length 
$1.4$~a.u.) with all hydrogen atoms placed co-linearly, hence the general formula $(\mbox{H}_2)_n$. The distance between 
centers of mass of two neighboring hydrogen molecules is equal $4.2$~a.u. This system behaves as an insulator even in 
the limit of an infinite chain length, i.e. $n\rightarrow\infty$, which makes the single-reference CC approach valid.

We performed calculations of the quadruples corrections using the exact CCSDT(Q) method as implemented in \textsc{CFour} 
and the $\mathcal{L}_{\rm (Q)}$ functional for the $(\mbox{H}_2)_n$ system, $n=1,2,\ldots,10$, within the 
cc-pVDZ basis set. We focus solely on the scaling of the $N_{\rm qua}$ parameter and hence employ the full space of 
triple excitations in the SVD-CCSDT calculations preceding the evaluation of the $\mathcal{L}_{\rm (Q)}$ 
functional. In Fig.~\ref{fig:error-scaling} we show relative errors in the calculated (Q) correction for a 
representative value of the quadruple excitation subspace size equal to $N_{\rm qua}=10\cdot n$ and $N_{\rm 
qua}=15\cdot n$, where $n$ is the chain length. As the value of $n$ increases, the relative error quickly reaches the 
asymptotic values of approximately 11\% and 1\%, respectively. In the region beyond $n\approx 7$ the error is 
essentially constant with only minor fluctuations of the order of 0.1\% or less. This confirms that in order to 
maintain a constant relative accuracy in the (Q) correction it is sufficient to make $N_{\rm qua}$ proportional to the 
system size.

\begin{figure}[ht]
\includegraphics[scale=0.75]{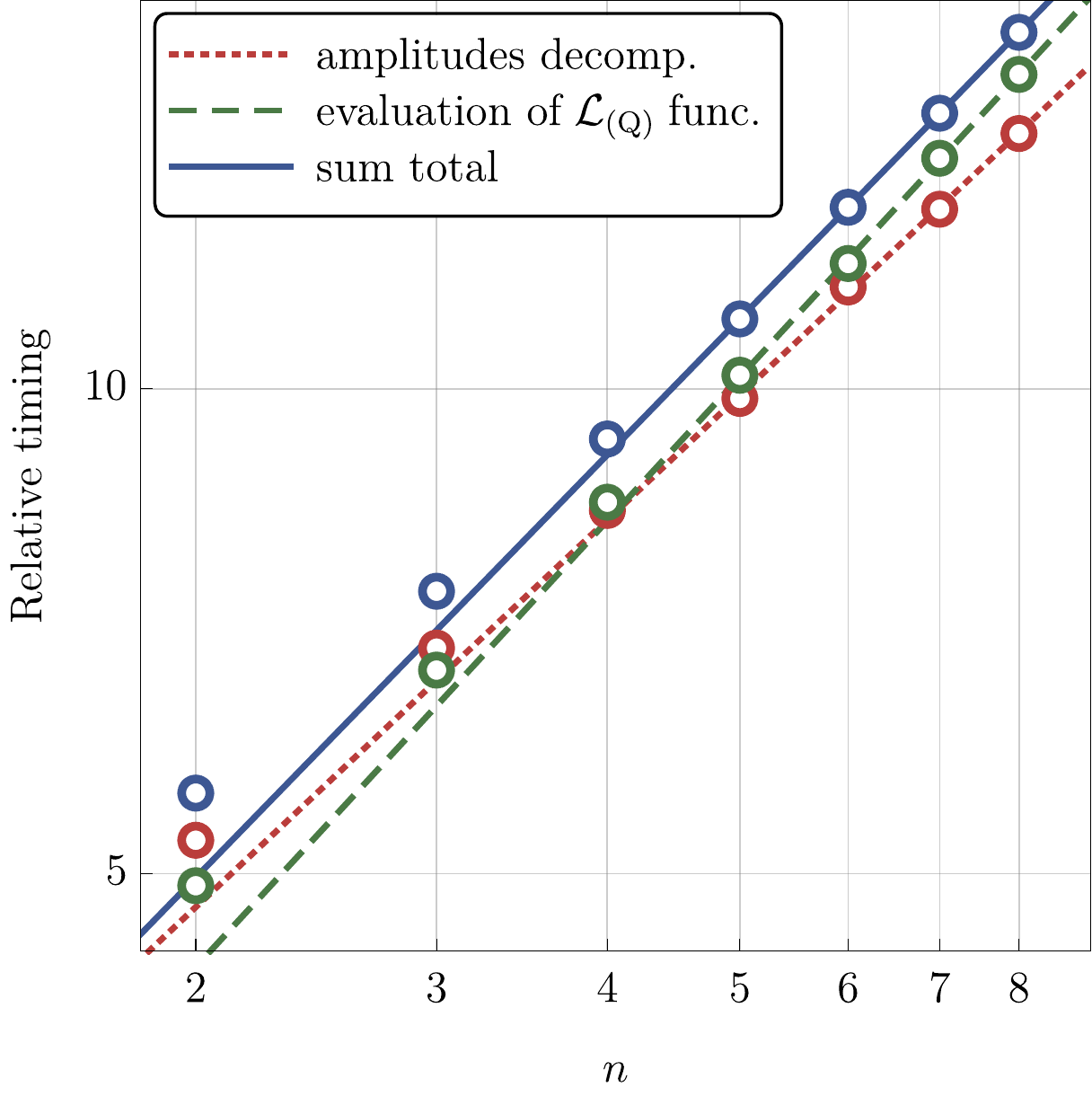}
\caption{\label{fig:time-scaling}
Relative timings of two major computational steps of the proposed formalism (cc-pVDZ basis set) for linear alkanes (C$_n$H$_{2n+2}$) as a function of the chain length, $n$. Logarithmic scale is used on both axes.
}
\end{figure}

Another numerical demonstration necessary to confirm the theoretical findings of the present work is related to the computational cost of evaluating the $\mathcal{L}_{\rm (Q)}$ functional. The analysis of the working equations of the proposed method given in Secs.~\ref{subsec:nonitt4}-\ref{subsec:quadratic2} (and in Supporting Information) led to the conclusion that all terms in the $\mathcal{L}_{\rm (Q)}$ functional can be computed with the cost proportional to $N^7$ in the rate-determining steps. Regarding the decomposition of the $T_4$ and $L_4$ amplitudes, they formally share the same $N^7$ asymptotic cost, but involve only one $N^7$ step that has a small prefactor. Therefore, we expect that this component of the method would possess, in practice, a cost proportional to $N^6$. Finally, the decomposition of the $L_3$ amplitudes was found to scale as $N^6$.

To confirm the aforementioned findings numerically, we perform calculations for model systems that can be systematically increased in size. In contrast with the calculations considered at the beginning of this section, the reference CCSDT(Q) calculations are not involved here and hence it is feasible to study a more chemically appealing model system of linear alkanes, C$_n$H$_{2n+2}$. Molecular geometries were taken from Ref.~\onlinecite{lesiuk20}. As an illustrative example, we set $\nsvd=N_{\mathrm{MO}}$ and $\nqua=N_{\mathrm{MO}}$, where $N_{\mathrm{MO}}$ is the number of orbitals in the system, so that both parameters increase linearly with the chain length. The core $1s^2$ orbitals of the carbon atoms are not correlated. In Fig.~\ref{fig:time-scaling} we report total timings of the calculations, as well timings for two major parts, namely (i) evaluation of the $\mathcal{L}_{\rm (Q)}$ functional, and (ii) decomposition of the $T_4$, $L_3$, and $L_4$ amplitudes. For clarity, the timings are given in relation to the calculations for methane. To confirm that the results given in Fig.~\ref{fig:time-scaling} match the theoretical predictions, we fitted the timings with the functional form $a\cdot n^b$ (a linear function on a logarithmic scale) for $n=3-8$. We obtained the exponents $b=6.64$ and $b=5.76$ for the parts (i) and (ii), respectively. The empirically found values of the exponents are in both cases somewhat smaller than predicted theoretically ($7$ and $6$). This can be explained by the fact that both parts of the calculations involve also many lower-scaling steps such as computation of intermediate quantities, etc. While the cost of such steps is asymptotically marginal, they still contribute non-negligibly to the total workload for systems that can be studied at present.

\subsection{Calibration of the method}
\label{subsec:param}

In this section we study errors of the proposed formalism in reproduction of the absolute (Q) correction. For this purpose we selected a set of $16$ small molecules comprising $2-5$ first row atoms. Within the cc-pVTZ basis set employed in the calculations, the largest molecule is described by $118$ atomic orbitals and hence the conventional CCSDT(Q) calculations are feasible with a reasonable computational cost. Therefore, the values of the exact (Q) correction are available for each molecule and shall be used as reference in the present section. The list of molecular systems used in the benchmark calculations and their structures in Cartesian coordinates are provided in the Supporting Information. To assure that the error resulting from the density-fitting approximation does not contaminate the final conclusions of this section, a large cc-pV5Z-RI auxiliary basis set is used. We verified that this leads to errors of no larger than a few $\mu$H in the correlation energies which is entirely negligible in the present context.

An important aspect of the analysis provided below relates to the determination of the recommended values of the quantities $\nsvd$ and $\nqua$ that serve as parameters in the rank-reduced formalism in the present work. The former parameter determines the size of the triple excitation subspaces used in $T_3$ and $L_3$ operators, see Eq. (\ref{t3}), while the latter serves the same purpose in the case of the quadruple excitation subspaces in $T_4$ and $L_4$. As the cost of the calculations increases steeply with increasing $\nsvd$ and $\nqua$, it is necessary to recommend a way of determining these parameters for a given system such that a sufficient level of accuracy is attained while the computational cost is simultaneously minimized. Since both $\nsvd$ and $\nqua$ increase linearly with the system size, it is convenient to tie them to some quantity that shares the same property, but is known upfront for a given system. In this way, the parameters can be easily transferred between molecules of different size. Similarly to previous works, we express the parameters $\nsvd$ and $\nqua$ as a fraction times the total number of active molecular orbitals in the system, $N_{\mathrm{MO}}$ (frozen-core orbitals and possibly frozen virtual orbitals are excluded). In other words, these parameters are given by $\nsvd=x\cdot N_{\mathrm{MO}}$ and $\nqua=y\cdot N_{\mathrm{MO}}$, where $x$, $y$ are asymptotically independent of the system size. Note that both $x$ and $y$ may be larger than the unity. It has been shown in Ref.~\onlinecite{lesiuk20} that in reproduction of the CCSDT correlation energy $\nsvd\approx N_{\mathrm{MO}}$ is sufficient in usual applications. However, it cannot be guaranteed \emph{a priori} that a similar size of the triple excitation subspace is adequate in determination of the (Q) corrections.

\begin{figure}[ht]
\includegraphics[scale=1.0]{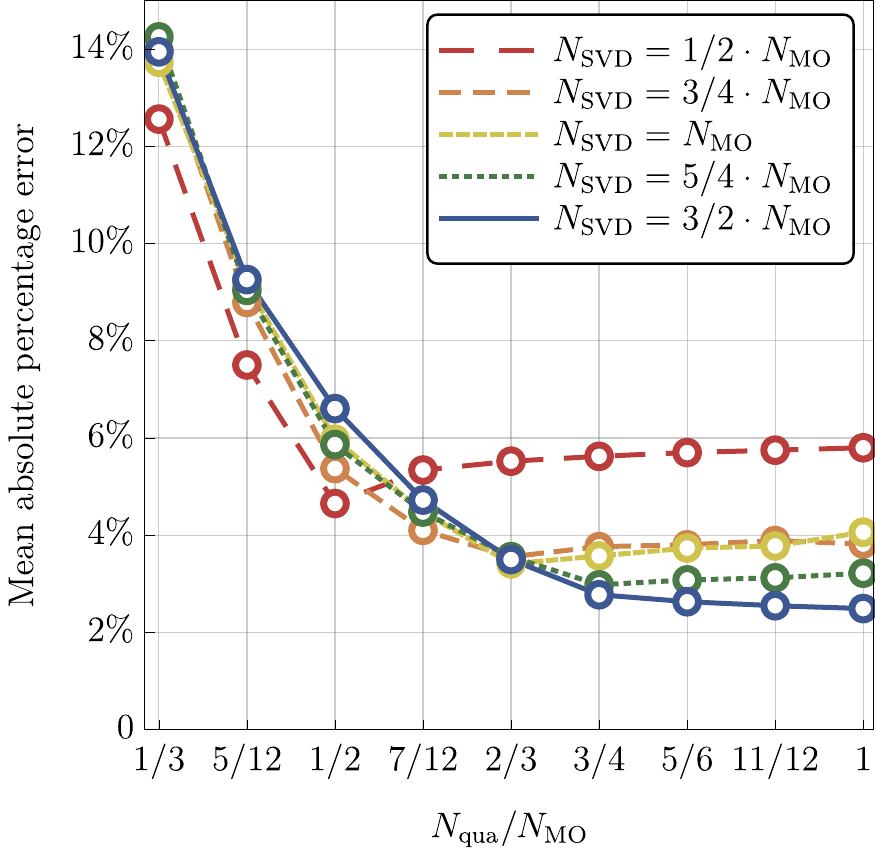}
\caption{\label{fig:total1}
Mean absolute percentage error in the (Q) correction obtained using the $\mathcal{L}_{\rm (Q)}$ functional 
(cc-pVTZ basis set) as a function of $\nqua$ for several representative values of the $\nsvd$ parameter, see the 
legend. The exact CCSDT(Q) results are used as a reference. The symbol $N_{\mathrm{MO}}$ denotes the total number of 
orbitals in a given system.
}
\end{figure}

Since the molecules included in the test set vary considerably in size, it is necessary to use a size-intensive error measure to compare the results. First, we consider the mean absolute percentage error (MAPE) averaged over all molecules. In Fig.~\ref{fig:total1} we plot MAPE as a function of $\nqua$ varied from $\frac{1}{3}N_{\mathrm{MO}}$ to $N_{\mathrm{MO}}$ for several representative values of $\nsvd$, namely $\nsvd=x\cdot N_{\mathrm{MO}}$ with $x=\frac{1}{2},\frac{3}{4},1,\frac{5}{4},\frac{3}{2}$. Overall, for each value of $\nsvd$ individually, we observe a similar trend in error decay as $\nqua$ is increased. Initially, the error vanishes rapidly, which is followed by a plateau region where the error stabilizes, see Fig.~\ref{fig:total1}. Beyond this point, further increase of the $\nqua$ parameters leads to no appreciable accuracy improvements, and in some cases the error even increases by a tiny  amount. Clearly, in this region the accuracy is limited by the error of the triply-, and not quadruply-excited, amplitudes. This is further confirmed by the observation that the error in the plateau region depends significantly on $\nsvd$. For $\nsvd=\frac{1}{2}N_{\mathrm{MO}}$ the error stabilizes at the level of around 6\%; this decreases to around 4\% for $\nsvd=N_{\mathrm{MO}}$, and to slightly above 2\% for $\nsvd=\frac{3}{2}N_{\mathrm{MO}}$. 

To recommend values of the parameters $\nsvd$ and $\nqua$ that shall be used in future calculations, we require that the 
MAPE in Fig.~\ref{fig:total1} should be at the order of a few percent, according to the discussion in the Introduction. 
The smallest triple excitation subspace that systematically delivers the accuracy better than 5\% corresponds to 
$\nsvd=\frac{3}{4}N_{\mathrm{MO}}$ or $\nsvd=N_{\mathrm{MO}}$. Smaller values of $\nsvd$ are not recommended, unless 
supported by some reference calculations that confirm their reliability or if larger errors are acceptable. Further 
increase, to about $\nsvd=\frac{3}{2}N_{\mathrm{MO}}$, of the parameter $\nsvd$ is necessary to reach accuracy levels of 
2\% or so. Regarding the value of the second parameter, it is desirable to set $\nqua$ to a value that (for a given 
$\nsvd$) corresponds as closely as possible to the onset of the plateau region. In this way the computational cost of 
the procedure is minimized without compromising the accuracy. A conservative choice is to set $\nqua=\frac{2}{3}\nsvd$, 
such that for each $\nsvd\geq N_{\mathrm{MO}}$ the value of $\nqua$ lies well-within the plateau region. We recommend 
this setup in future calculations. This approach has one additional advantage: the values of $\nsvd$ and $\nqua$ do not 
have to be varied independently. Instead, they are increased simultaneously which makes the results easier to analyze 
and represent.

\begin{figure}[ht]
\includegraphics[scale=0.75]{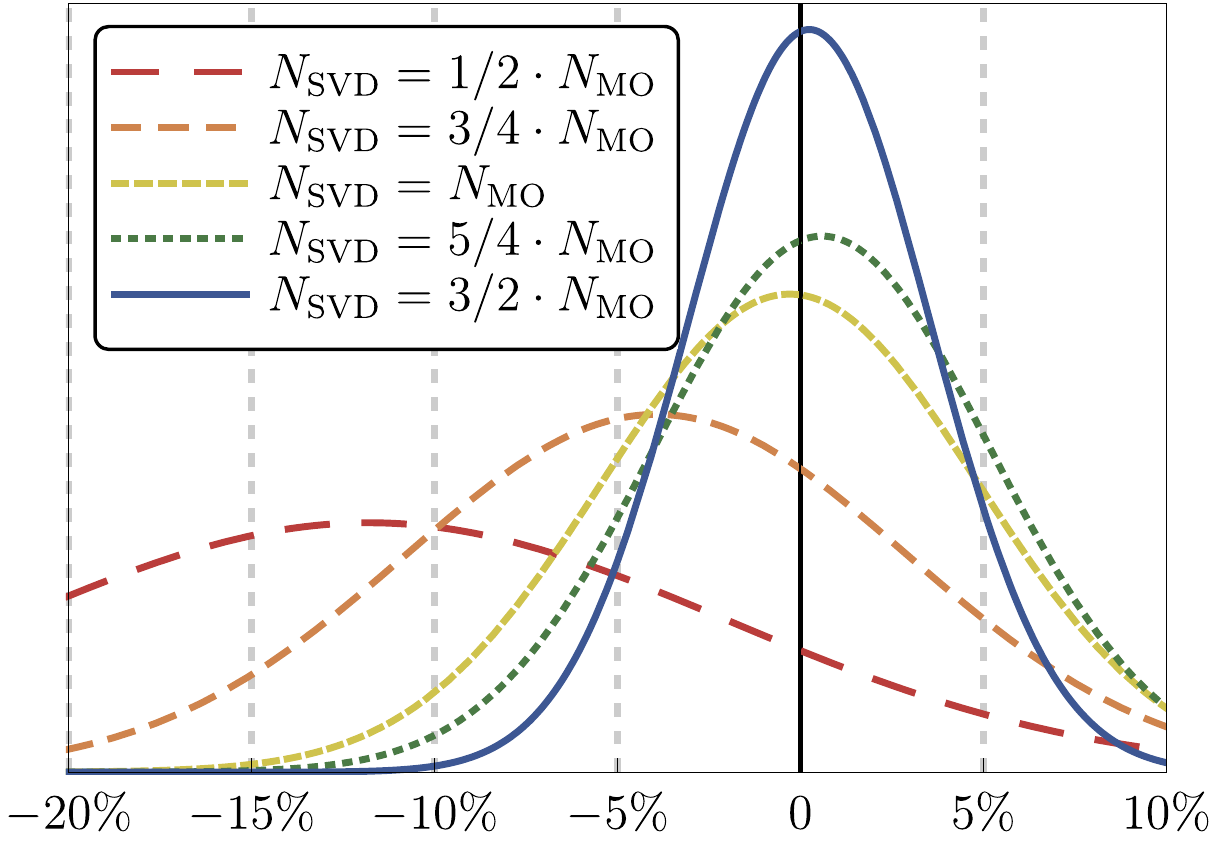}
\caption{\label{fig:total2}
Distribution of relative errors (in percent) in the (Q) correction obtained using the $\mathcal{L}_{\rm (Q)}$ functional (cc-pVTZ basis set) for several representative values of the $\nsvd$ parameter, see the legend. For each value of $\nsvd$, the parameter $\nqua$ is set to $\frac{2}{3}\nsvd$. The exact CCSDT(Q) results are used as a reference. The symbol $N_{\mathrm{MO}}$ denotes the total number of orbitals in a given system.
}
\end{figure}

Thus far we have concentrated on MAPE as a measure of error of the proposed formalism. While this error measure is particularly important from the practical point of view, it provides little information about the error distribution and its characteristics. To fill this gap we consider signed percentage error defined as
\begin{align}
\label{errort}
 \Delta^{(i)} = 100\%\cdot \frac{\mathcal{L}_{\rm (Q)}^{(i)}-E_{\mathrm{(Q)}}^{(i)}}{E_{\mathrm{(Q)}}^{(i)}},
\end{align}
where the index $i$ enumerates the molecules in the test set, and $E_{\mathrm{(Q)}}^{(i)}$ is the reference (exact) value of the (Q) correction for the $i$-th molecule. For the purposes of statistical analysis we calculate the mean error, $\bar{\Delta}$, and its standard deviation, $\Delta_{\mathrm{std}}^2$ using the well-known formulas. For all cases considered here we found that the error distribution is normal to a good degree of approximation. Therefore, for clarity we represent the 
error measures $\bar{\Delta}$ and $\Delta_{\mathrm{std}}$ graphically in Fig.~\ref{fig:total2} in terms of normalized Gaussian distributions. As illustrative examples we consider $\nsvd=x\cdot N_{\mathrm{MO}}$ 
with $x=\frac{1}{2},\frac{3}{4},1,\frac{5}{4},\frac{3}{2}$, and in each case we set $\nqua=\frac{2}{3}\nsvd$ according 
to the discussion above. As seen in Fig.~\ref{fig:total2}, the accuracy obtained with $\nsvd=\frac{1}{2}\cdot 
N_{\mathrm{MO}}$ and $\nsvd=\frac{3}{4}\cdot N_{\mathrm{MO}}$ is not satisfactory. Although in the latter case the 
mean error is acceptable ($\bar{\Delta}<5\%$), the corresponding standard deviation is still large 
($\Delta_{\mathrm{std}}\approx 6.8\%$) and hence the error distribution is rather broad. The results are improved 
considerably for $\nsvd=N_{\mathrm{MO}}$ and $\nsvd=\frac{5}{4} \cdot N_{\mathrm{MO}}$, where the mean error 
decreases to $\bar{\Delta}\approx -0.29\%$ and $\bar{\Delta}\approx 0.59\%$, respectively. This is accompanied by a 
significant reduction of the standard deviation. Finally, by increasing the value of the parameter $\nsvd$ to 
$\frac{3}{2}\cdot N_{\mathrm{MO}}$, the mean error is further reduced slightly ($\bar{\Delta}\approx 0.24\%$), 
similarly as the standard deviation ($\Delta_{\mathrm{std}}\approx 3.3\%$).

Next, we consider errors of the proposed formalism in reproduction of relative energies. To this end, we prepared the following set of $10$ chemical reactions:
{\small
\begin{enumerate}
\setlength\itemsep{-0.5em}
\item F$_2$ + H$_2$ $\rightarrow$ 2HF,
\item H$_2$O$_2$ + H$_2$ $\rightarrow$ 2H$_2$O,
\item CO + H$_2$ $\rightarrow$ H$_2$CO,
\item CO + 3H$_2$ $\rightarrow$ CH$_4$ + H$_2$O,
\item N$_2$ + 3H$_2$ $\rightarrow$ 2NH$_3$,
\item HCOOH $\rightarrow$ CO$_2$ + H$_2$,
\item CO + H$_2$O $\rightarrow$ CO$_2$ + H$_2$,
\item HCN + H$_2$O $\rightarrow$ CO + NH$_3$,
\item HCN + NH$_3$ $\rightarrow$ N$_2$ + CH$_4$,
\item H$_2$CO + H$_2$O$_2$ $\rightarrow$ HCOOH + H$_2$O.
\end{enumerate}
}
Contribution of the (Q) correction to each reaction energy has been computed with three different schemes (all within the cc-pVTZ orbital basis set). First, the exact CCSDT(Q) calculations that employ neither decomposition of the excitation amplitudes nor density-fitting approximation are used as a benchmark. The second and third scheme involves the quadratic (Q) functional with the recommended settings, namely $\nqua=\frac{2}{3}\nsvd$, and for several representative $\nsvd$. In the second scheme we use a large cc-pV5Z-RI auxiliary basis set for density-fitting approximation and hence the DF errors are marginal. In the third scheme the same settings are used, except that the standard cc-pVTZ-RI auxiliary basis set (matched to the used orbital cc-pVTZ) is employed. This enables us to study the impact of the density-fitting approximation on the quality of the results and establish whether the conventional auxiliary basis sets available in the literature are sufficient for the present purposes.

{\setlength{\tabcolsep}{10pt}
\begin{table}[ht]
 \caption{Contribution of the (Q) correction to the reaction energies calculated using the exact CCSDT(Q) method and using the quadratic (Q) functional as as function of $\nsvd$ (with fixed $\nqua=\frac{2}{3}\nsvd$). The mean absolute error (MAE) and the standard deviation of the error (STD) are given in the last two rows. The cc-pVTZ orbital basis set is used together with a large cc-pV5Z-RI auxiliary basis set to eliminate the error due to the density-fitting approximation. All values are given in kJ/mol.}
\label{tab:reactions1}
\begin{tabular}{lcccccc}
\hline\hline
 & & \multicolumn{5}{c}{$N_{\mathrm{SVD}}$\textsuperscript{a}} \\
reaction & $\;\;\;$exact
         & $\frac{1}{2}N_{\mathrm{MO}}$
         & $\frac{3}{4}N_{\mathrm{MO}}$
         & $N_{\mathrm{MO}}$
         & $\frac{4}{3}N_{\mathrm{MO}}$
         & $\frac{3}{2}N_{\mathrm{MO}}$ \\
\hline
1  & \phantom{$-$}2.88 & $-$0.12 & $-$0.10 & $-$0.12 & $-$0.08 & $-$0.06 \\ 
2  & \phantom{$-$}1.52 & $-$0.14 & $-$0.20 & $-$0.18 & $-$0.10 & $-$0.07 \\ 
3  & \phantom{$-$}0.29 & $-$0.12 & \phantom{$-$}0.02 & \phantom{$-$}0.06 & 
\phantom{$-$}0.02 & \phantom{$-$}0.02 \\ 
4  & \phantom{$-$}1.61 & $-$0.26 & $-$0.10 & \phantom{$-$}0.04 & \phantom{$-$}0.01 & \phantom{$-$}0.01 \\ 
5  & \phantom{$-$}3.15 & $-$0.17 & $-$0.06 & \phantom{$-$}0.12 & \phantom{$-$}0.13 & \phantom{$-$}0.10 \\ 
6  & $-$1.06 & \phantom{$-$}0.36 & \phantom{$-$}0.01 & $-$0.17 & $-$0.13 & $-$0.05 \\ 
7  & $-$1.33 & \phantom{$-$}0.50 & \phantom{$-$}0.29 & \phantom{$-$}0.12 & \phantom{$-$}0.03 & \phantom{$-$}0.01 \\ 
8  & \phantom{$-$}1.63 & \phantom{$-$}0.34 & \phantom{$-$}0.28 & \phantom{$-$}0.09 & \phantom{$-$}0.20 & \phantom{$-$}0.08 \\ 
9  & \phantom{$-$}0.08 & \phantom{$-$}0.25 & \phantom{$-$}0.25 & \phantom{$-$}0.00 & \phantom{$-$}0.08 & \phantom{$-$}0.03 \\ 
10 & \phantom{$-$}1.06 & \phantom{$-$}0.12 & \phantom{$-$}0.06 & \phantom{$-$}0.05 & \phantom{$-$}0.04 & \phantom{$-$}0.02 \\  
\hline
MAE & $\;\;\;-$ & \phantom{$-$}0.24 & \phantom{$-$}0.14 & \phantom{$-$}0.10 & \phantom{$-$}0.08 & \phantom{$-$}0.07 \\
STD & $\;\;\;-$ & \phantom{$-$}0.27 & \phantom{$-$}0.17 & \phantom{$-$}0.12 & \phantom{$-$}0.10 & \phantom{$-$}0.07 \\
\hline\hline
\end{tabular}
\vspace{-0.25cm}
\begin{flushleft}
\small \textsuperscript{a} Errors (for a given $N_{\mathrm{SVD}}$) with respect to the exact result.
\end{flushleft}
\end{table}
}

In Tables~\ref{tab:reactions1}~and~\ref{tab:reactions2} we report results of the calculations with the large and standard density-fitting basis sets, respectively. As expected, for $\nsvd=\frac{1}{2}N_{\mathrm{MO}}$ and $\nsvd=\frac{3}{4}N_{\mathrm{MO}}$ average errors are substantial from the present point of view. However, as the size of the excitation subspace is increased to $\nsvd=N_{\mathrm{MO}}$, the average absolute errors decrease to about $0.1$~kJ/mol. This confirms that the recommended settings perform well in determination of the relative energies. Additionally, by comparing the results presented in Tables~\ref{tab:reactions1}~and~\ref{tab:reactions2} one finds that the density-fitting approximation has a tiny impact on the quality of the results. The average DF errors are of the order of $0.01$~kJ/mol and even the largest error found for the test set is well below $0.1$~kJ/mol. Therefore, we conclude that the standard auxiliary basis sets (optimized for a given orbital basis) are sufficient for accurate determination of the (Q) correction. 

{\setlength{\tabcolsep}{10pt}
\begin{table}[ht]
 \caption{The same data as in Table~\ref{tab:reactions1} but obtained using the standard cc-pVTZ-RI auxiliary basis set for the density-fitting approximation.}
\label{tab:reactions2}
\begin{tabular}{lcccccc}
\hline\hline
 & & \multicolumn{5}{c}{$N_{\mathrm{SVD}}$\textsuperscript{a}} \\
reaction & $\;\;\;$exact
         & $\frac{1}{2}N_{\mathrm{MO}}$
         & $\frac{3}{4}N_{\mathrm{MO}}$
         & $N_{\mathrm{MO}}$
         & $\frac{4}{3}N_{\mathrm{MO}}$
         & $\frac{3}{2}N_{\mathrm{MO}}$ \\
\hline
1  & \phantom{$-$}2.88 & $-$0.16 & $-$0.09 & $-$0.06 & $-$0.12 & $-$0.06 \\ 
2  & \phantom{$-$}1.52 & $-$0.14 & $-$0.29 & $-$0.18 & $-$0.11 & $-$0.07 \\ 
3  & \phantom{$-$}0.29 & $-$0.11 & \phantom{$-$}0.02 & \phantom{$-$}0.08 & \phantom{$-$}0.03 & \phantom{$-$}0.03 \\ 
4  & \phantom{$-$}1.61 & $-$0.25 & $-$0.14 & \phantom{$-$}0.05 & \phantom{$-$}0.01 & \phantom{$-$}0.09 \\ 
5  & \phantom{$-$}3.15 & $-$0.18 & $-$0.08 & \phantom{$-$}0.13 & \phantom{$-$}0.13 & \phantom{$-$}0.10 \\ 
6  & $-$1.06 & \phantom{$-$}0.24 & $-$0.17 & $-$0.17 & $-$0.12 & $-$0.05 \\ 
7  & $-$1.33 & \phantom{$-$}0.39 & $-$0.06 & \phantom{$-$}0.14 & \phantom{$-$}0.04 & \phantom{$-$}0.03 \\ 
8  & \phantom{$-$}1.63 & \phantom{$-$}0.31 & \phantom{$-$}0.29 & \phantom{$-$}0.07 & \phantom{$-$}0.20 & \phantom{$-$}0.08 \\ 
9  & \phantom{$-$}0.08 & \phantom{$-$}0.25 & \phantom{$-$}0.22 & \phantom{$-$}0.00 & \phantom{$-$}0.07 & \phantom{$-$}0.03 \\ 
10 & \phantom{$-$}1.06 & \phantom{$-$}0.12 & $-$0.20 & \phantom{$-$}0.04 & \phantom{$-$}0.03 & \phantom{$-$}0.02 \\ 
\hline
MAE & $\;\;\;-$ & \phantom{$-$}0.21 & \phantom{$-$}0.16 & \phantom{$-$}0.09 & \phantom{$-$}0.08 & \phantom{$-$}0.07 \\
STD & $\;\;\;-$ & \phantom{$-$}0.24 & \phantom{$-$}0.18 & \phantom{$-$}0.11 & \phantom{$-$}0.11 & \phantom{$-$}0.07 \\
\hline\hline
\end{tabular}
\vspace{-0.25cm}
\begin{flushleft}
\small \textsuperscript{a} Errors (for a given $N_{\mathrm{SVD}}$) with respect to the exact result.
\end{flushleft}
\end{table}
}

It must be pointed out that with the parameters $\nsvd$ and $\nqua$ set as fixed multiples of some system-dependent quantity, the proposed method is not guaranteed to be size-consistent. This is true even one selects a multiple of a quantity that scales linearly with the system size, such as $N_{\mathrm{MO}}$ suggested in the previous paragraphs. While the results presented in this work, e.g. for the reaction energies, show that the size-inconsistency error is tiny, this can still become problematic in applications, e.g. to weakly-interacting systems, where proper cancellation of non-physical size-inconsistent contributions is important. One may argue from a pragmatic standpoint that if significant size-inconsistency errors are encountered, the simplest remedy is to increase the constant factor that relates $\nsvd$ and $\nqua$ to $N_{\mathrm{MO}}$. As the rank-reduced formalism is built upon CC methods which are rigorously size-extensive, this approach is, in principle, always able to decrease these errors to acceptable levels. However, this line of reasoning is not fully satisfactory as it may lead to a significant increase of the computational costs. A more suitable approach would be to determine the size of the excitation subspace adaptively for a given molecule based on some numerical threshold that is transferable between systems. However, this problem is non-trivial as the higher-order orthogonal iteration algorithm applied to higher-order amplitudes does not provide a natural numerical parameter that can be used for the truncation, in contrast to, e.g. the diagonalization approach adopted in Ref.~\onlinecite{parrish19} for the doubly-excited amplitudes. Our preliminary tests have shown that a simple threshold on, e.g. the eigenvalues of the $M_{ai,bj}$ matrix defined by Eq.~(\ref{mmatrix}) are not entirely satisfactory. Therefore, a more elaborate scheme is required which employs, e.g. a threshold on the cumulative eigenvalues of the $M_{ai,bj}$ matrix. A complete analysis of this problem requires numerous benchmarks calculations analogous to the data presented in Ref.~\onlinecite{nagy21}. This is beyond the scope of the present paper and requires a separate study which is currently in progress.

Finally, after establishing the computational protocol that shall be used in subsequent applications, we can assess the computational performance of the proposed theory in comparison with the conventional (Q) implementation. While we have shown in the previous section that the rank-reduced formalism is characterized by a lower scaling with the system size than the exact (Q) method ($N^7$ vs. $N^9$), it is not yet clear how this translates into computational advantages. In fact, due to the overhead related to determination of the excitation subspace (and other steps of the rank-reduced calculations), one expects that the prefactor of the conventional (Q) algorithm is lower and hence there is a break-even point beyond which our method is favored. In order to approximately locate this break-even point, we compared the timings of the calculations reported in this section with the conventional (Q) algorithm as implemented in the \textsc{CFour} program (to allow a fair comparison, the same machine was used for both calculations and serial program execution was requested). Depending on the cardinal number of the basis set, the proposed algorithm becomes beneficial for systems with more than $100-150$ basis set functions. For example, for the formic acid molecule in the cc-pVTZ basis set ($115$ active orbitals), the rank-reduced calculations are three times faster than the conventional algorithm. In general, our tests revealed that the break-even point occurs faster for smaller basis sets. However, we also would like to point out that the current pilot implementation of the rank-reduced formalism can be improved by more extensive code optimization. While, in principle, the same is true for the conventional (Q) algorithm, implementations of the latter are much more mature.

\subsection{Isomerization energy of \emph{ortho}/\emph{meta} benzyne}
\label{subsec:application1}

As the first application of the proposed theory we study the isomers of benzyne molecule (C$_6$H$_4$) in singlet spin state. We are interested in the isomerization energy between \emph{ortho}- and \emph{meta}-benzyne. Benzynes have attracted a significant interest in recent years, both from experimental~\cite{jones72,wenthold98} and theoretical point of view~\cite{lindh95,cramer97,lindh99,moskaleva99,crawford01,smith05,karton09,ghigo14}, due to their unique electronic structure and chemical properties. In synthetic organic chemistry, \emph{ortho}-benzyne is a crucial intermediate in several important types of reactions, see the recent review paper of Tadross and Stoltz~\cite{tadross12} for an in-depth discussion. Benzynes are also found~\cite{hirsch18} to be the key 
intermediate in formation of polycyclic aromatic hydrocarbons -- carcinogenic and environmentally-harmful 
compounds. In particular, the \emph{ortho}- and \emph{meta}-benzyne isomerization has been proposed to constitute an important step of various fragmentation and decomposition reaction pathways~\cite{crews73,matsugi12,martinez16}. From the theoretical standpoint, benzynes are known to possess a singlet diradical character of the ground-state energy level and it has been shown that static correlation effects play an important role in description of these systems. Therefore, benzynes are frequently employed in benchmark studies of novel quantum chemistry methods where accurate and reliable reference data is valuable.

Throughout the present section we adopt the convention that the isomerization energy $\Delta E$ is defined as
\begin{align}
 \Delta E = E_{\mathrm{\emph{meta}}} - E_{\mathrm{\emph{ortho}}}.
\end{align}
It is known that in the ground electronic singlet state the \emph{ortho} isomer is more stable and hence the total isomerization energy defined above is positive. However, individual contributions to the isomerization energy representing various physical contributions may be of an arbitrary sign. For clarity, positive contributions are understood to favor the \emph{ortho} isomer, while negative contributions -- the \emph{meta} isomer.

Besides studying the performance of the rank-reduced formalism for the isomers of benzyne, our goal is to show how the proposed method can be incorporated in the so-called composite electronic structure schemes in order to increase their accuracy, computational performance or range of applicability. Several families of composite schemes were proposed in the literature, e.g. Gaussian‐$n$ (G-$n$) originally introduced by Pople \emph{et al.}~\cite{curtiss90,curtiss91,curtiss98,curtiss07}, Weizmann-$n$ (W-$n$) model chemistry developed by Martin and collaborators~\cite{martin99,boese04,karton06}, or HEATxyz protocol in its several variants~\cite{tajti04,bomble06,harding08}. Nowadays, composite schemes are an important tool in, e.g. \emph{ab initio} thermochemistry calculations or theoretical prediction of quantitative chemical kinetics. The main idea behind the composite schemes is to split, e.g. the total electronic energy of a molecule, into a sum of one (in some variants, two) major components supplemented by a series of smaller additive corrections. Depending on the desired level of accuracy, the number of corrections varies from just a few to a dozen or so in the most demanding situations. As the corrections are small in absolute terms compared to the major components, they can be calculated less accurately -- usually employing a smaller basis set or with some additional approximations. Below we introduce a composite method that targets the accuracy comparable to the exact CCSDT(Q) theory, and employs the rank-reduced approach to the calculation of the (Q) correction and the SVD-CCSDT+ method for determination of the triple excitation contributions. 

The molecular geometries of \emph{ortho}- and \emph{meta}-benzyne were taken from the paper of Karton~\emph{et al.}~\cite{karton09} where they were optimized using the frozen-core CCSD(T)/cc-pVQZ method. Our preliminary study has shown that these structures are well-converged with respect to the basis set size and the CC level. Therefore, all subsequent calculations of contributions to the isomerization energy were performed at fixed CCSD(T)/cc-pVQZ geometries, unless explicitly stated otherwise.

We begin by considering the Hartree-Fock contribution to the isomerization energy which was calculated using the cc-pV$X$Z basis sets with $X$=D,T,Q,5. As expected, it converges very fast to the basis set limit, with the value calculated using the cc-pVQZ basis set, $106.90$~kJ/mol, differs from the result obtained within the cc-pV5Z basis set, $106.97$~kJ/mol, by just 0.07~kJ/mol. To further minimize the basis set incompleteness error we perform three-parameter extrapolation using the exponential formula
\begin{align}
\label{extra_exp}
 E_X = E_\infty + A\,e^{-B X},
\end{align}
where $E_\infty$, $A$, $B$ are fitted to reproduce the results obtained within the largest three basis sets. This leads to the final result $106.99$~kJ/mol. Taking into account the rapid convergence of the results with the basis set size, it is reasonable to assume that the error of this quantity is no larger than $0.05$~kJ/mol.

Next, we move on to the valence CCSD contribution using the same basis sets as for the Hartree-Fock method. One obtains $-37.08$, $-34.90$, $-33.52$, $-32.99$~kJ/mol with the $X$=D,T,Q,5 basis sets, respectively. The convergence pattern to the basis set limit is systematic, but the residual basis set error is still relatively large. To eliminate a significant portion of this error we employ the two-parameter Riemann extrapolation formula proposed in Ref.~\onlinecite{lesiuk19b}:
\begin{align}
\label{extra_riemann}
 E_\infty = E_X + X^4 \bigg[\frac{\pi^4}{90} - \sum_{n=1}^X n^{-4} \bigg] \Big( E_X - E_{X-1} \Big).
\end{align}
This stencil is used for extrapolation of all correlation energies in the remainder of the present work. The extrapolated CCSD contribution using the $X=\mathrm{Q}(4),5$ basis sets reads $-32.32 \pm 0.34$~kJ/mol, where the error was conservatively estimated to be equal to the half of the difference between the extrapolated value and the result in largest basis set available. This approach was found to provide reliable and conservative error estimates for small molecular systems at the same level of theory~\cite{lesiuk19b,czachor20}.

\begin{figure}[ht]
\includegraphics[scale=0.75]{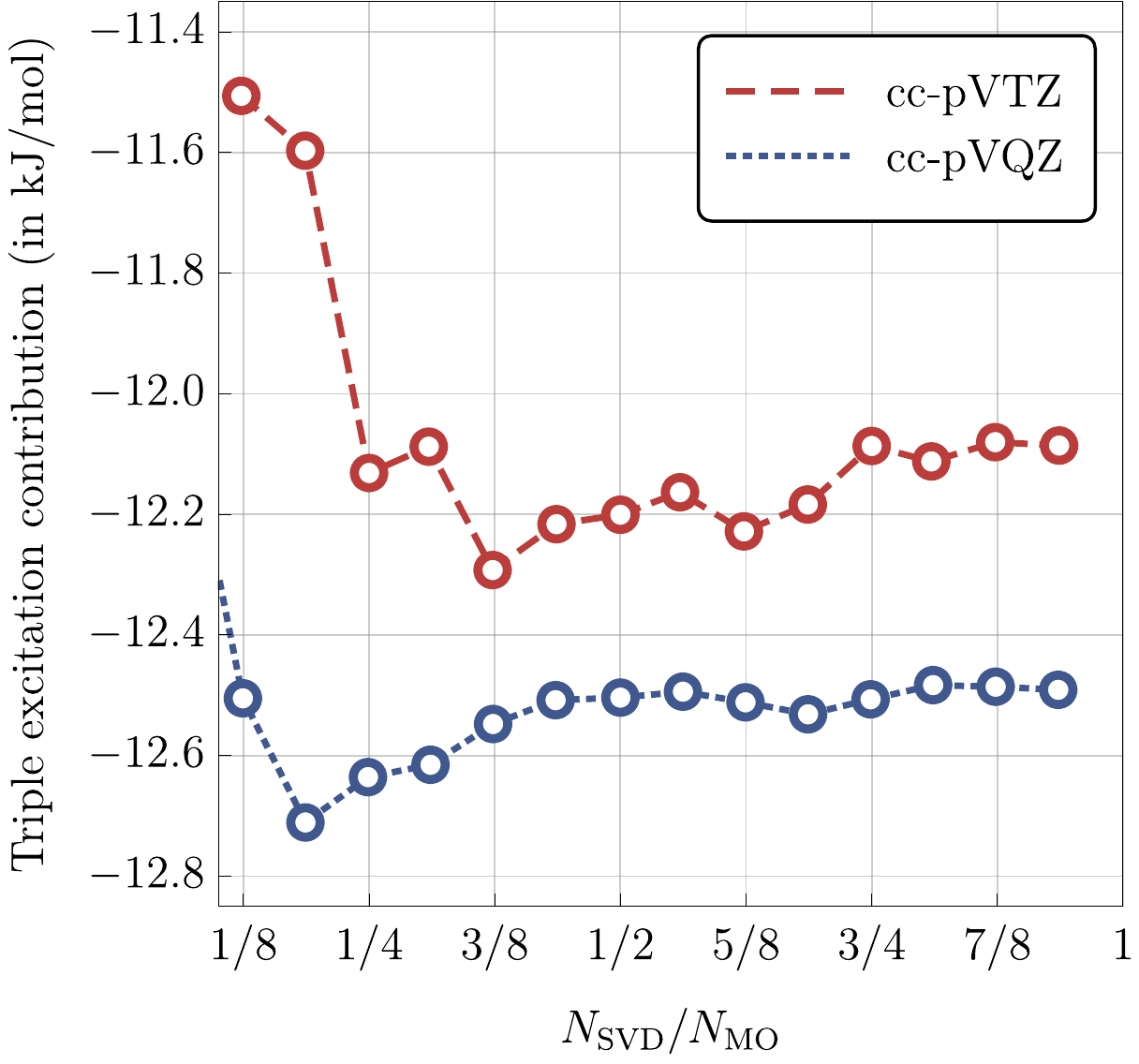}
\caption{\label{fig:t-benzyne}
Triple excitation contribution to the isomerization energy of \emph{ortho}/\emph{meta} benzyne calculated using the 
SVD-CCSDT+ method (cc-pVTZ and cc-pVQZ basis sets) as a function of the $\nsvd$ parameter. The symbol 
$N_{\mathrm{MO}}$ denotes the total number of orbitals in the system.
}
\end{figure}

The next important contribution to the isomerization energy is the effect of triple excitations, defined as the 
difference of the results obtained with the CCSDT and CCSD methods. Due to cost considerations, it is a common practice to 
split the triple excitations contribution into two parts, namely (i) the contribution of triple excitations captured by 
the CCSD(T) method and (ii) the remainder, i.e. the difference between the CCSD(T) and CCSDT results. This approach is 
justified by the fact that the first contribution is usually dominating and can be obtained with a 
larger basis set. However, in the present application we found this separation to be no longer beneficial if the 
SVD-CCSDT+ method is used for evaluation of the triple excitation effects. This is because the largest basis set we could 
use in both CCSD(T) and SVD-CCSDT+ calculations was cc-pVQZ. Despite significant effort, it was impossible to perform canonical
(T)/cc-pV5Z calculation using the hardware and software available to us, either due to excessive computational time or 
memory/disk space limitations. Taking into consideration that SVD-CCSDT+ is more accurate, we use it directly to 
determine the effects of triple excitations, bypassing the (T) method. In Fig.~\ref{fig:t-benzyne} we show the 
convergence of the $T_3$ contribution (cc-pVTZ and cc-pVQZ basis sets) to the isomerization energy as a function of the 
$N_{\mathrm{SVD}}$ parameter. Similarly as in the previous sections, this parameter is expressed as 
$N_{\mathrm{SVD}}=x\cdot N_{\mathrm{MO}}$, where $x\in(0,1]$. The results are remarkably stable with respect to the 
value of $N_{\mathrm{SVD}}$; for $x>\half$ the results change by less than $0.1$~kJ/mol with the smaller basis and 
$0.05$~kJ/mol with the larger basis. We take the results obtained with $x=1$ as the limit which leads to $12.12$~kJ/mol 
and $12.48$~kJ/mol within cc-pVTZ and cc-pVQZ basis sets, respectively. We assign the uncertainty of $0.05$~kJ/mol to 
both these values. To obtain the final value of the triple excitation contribution to the isomerization energy we 
perform two-point $X=\mathrm{T},\mathrm{Q}$ complete basis set extrapolation, giving $12.80\pm 0.17$~kJ/mol. Two sources 
of error contribute to the proposed uncertainty: the extrapolation error ($0.16$~kJ/mol) which was estimated in the same 
way as for the CCSD contribution, and the error due to the truncation of the triple excitation subspace ($0.05$~kJ/mol, 
see the discussion above related to the $N_{\mathrm{SVD}}$ parameter). Since both sources of error can be viewed as 
independent, the final error is calculated by summing their squares and taking the square root, according to the usual 
rules of error propagation.

\begin{figure}[ht]
\includegraphics[scale=0.75]{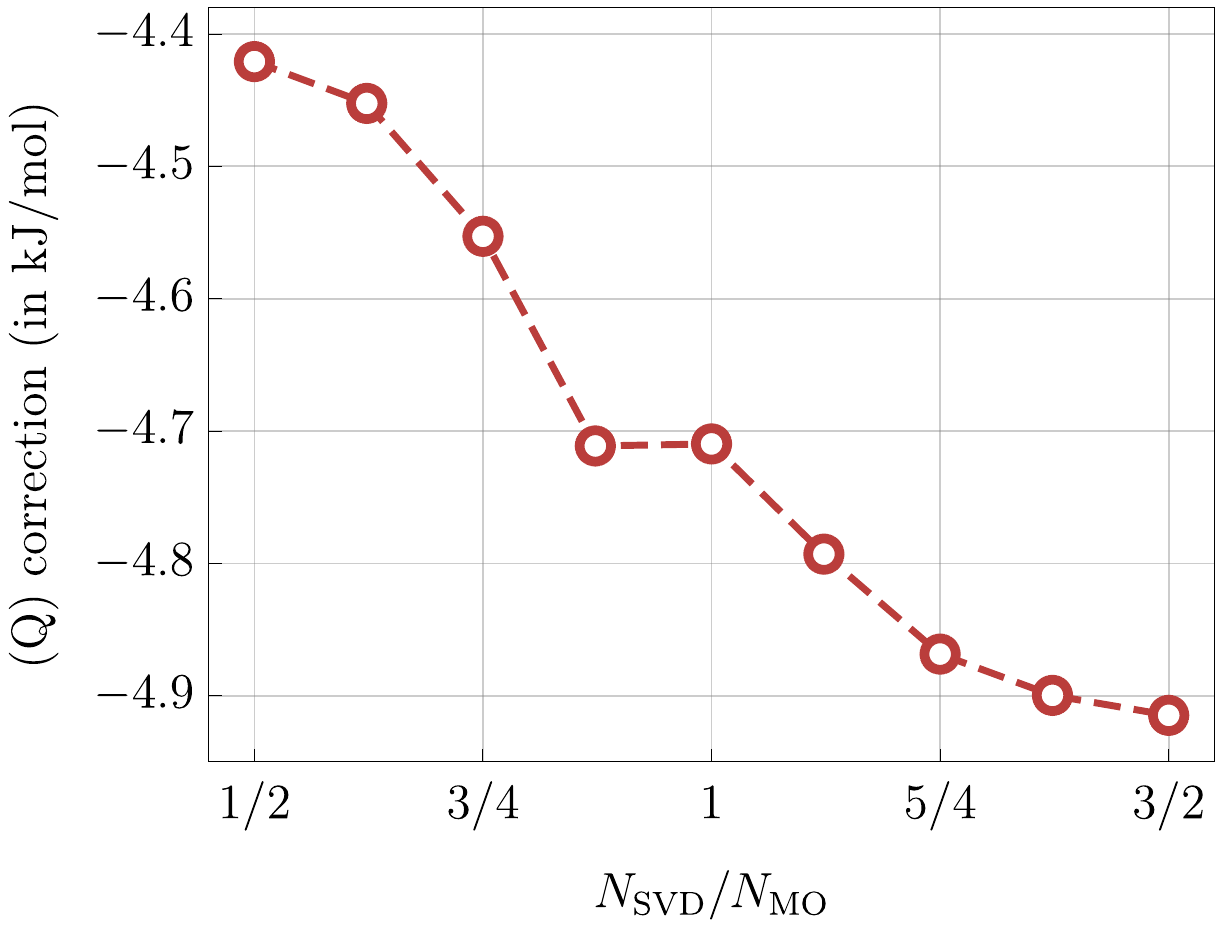}
\caption{\label{fig:q-benzyne}
Quadruple excitation contribution to the isomerization energy of \emph{ortho}/\emph{meta}-benzyne calculated using 
the $\mathcal{L}_{\rm (Q)}$ functional (cc-pVTZ basis set) as a function of the $\nsvd$ parameter. For each value of 
$\nsvd$, the parameter $\nqua$ is set to $\frac{2}{3}\nsvd$. The symbol $N_{\mathrm{MO}}$ denotes the total number 
of orbitals in the system.
}
\end{figure}

Finally, we move on to the calculation of the (Q) correction which accounts for the quadruple excitation effects. For this purpose we adopt the quadratic functional formalism introduced in the present work combined with the cc-pVTZ basis set. We also follow the recommendations stated in the previous section and set $\nqua=\frac{2}{3}\nsvd$ in all calculations. In Fig.~\ref{fig:q-benzyne} we present the (Q) contribution to the isomerization energy as a function of the $\nsvd$ parameter. Beyond $\nsvd=N_{\mathrm{MO}}$ changes in the (Q) correction are increasingly smaller. For example, upon increasing this parameter from $\nsvd=N_{\mathrm{MO}}$ to $\nsvd=\frac{5}{4}N_{\mathrm{MO}}$, the (Q) correction decreases by about $0.15$~kJ/mol, while further increase to $\nsvd=\frac{3}{2}N_{\mathrm{MO}}$ affects it only by ca. $0.04$~kJ/mol.
By following the trend seen in Fig.~\ref{fig:q-benzyne} one can expect that by further increase of the $\nsvd$ parameter the (Q) correction will still decrease slightly. However, the changes are expected to be insignificant; even assuming the worst case scenario that the convergence of the (Q) correction is inversely proportional to $\nsvd$, the limit would be less than $0.1$~kJ/mol away from the value obtained with $\nsvd=\frac{3}{2}N_{\mathrm{MO}}$. As a result, we assume that the (Q) correction to the isomerization energy is equal to the value obtained for $\nsvd=\frac{3}{2}N_{\mathrm{MO}}$, and assign conservative $0.1$~kJ/mol error bars, giving $-4.92\pm 0.10$~kJ/mol. We neglect the basis set incompleteness error in calculation of the (Q) correction. The computations of the (Q) correction using the quadratic functional for benzyne molecule with $\nsvd=\frac{3}{2}N_{\mathrm{MO}}$ and $\frac{2}{3}\nsvd$ (cc-pVTZ basis set) take about 2 days on 14 cores of AMD Opteron$^{\mathrm{TM}}$ Processor 6174.

The last major contribution to the isomerization energy is the zero-point vibration energy (ZPVE). Unfortunately, computation of this quantity at the CC level is costly, especially if a large basis set is required. For this reason, we employ the B3LYP/cc-pVTZ method to determine the ZPVE correction to the isomerization energy. Within the harmonic oscillator approximation the ZPVE contribution equals to $-4.34$~kJ/mol. This value is further scaled by the recommended factor $f=0.9764$ to take the anharmonic effects into account~\cite{sinha04}, giving $-4.24$~kJ/mol. In order to estimate the error of this quantity, we note that in a recent work B3LYP/cc-pVTZ method~\cite{vosko80,lee88,becke93,stephens94} was found to perform extremely well in comparison with CCSD(T) for molecules composed of first-row atoms~\cite{bakowies21}. This is especially true for hydrocarbons, where the average deviation from CCSD(T) is just about 1\%. Therefore, we conservatively assume that the error of the ZPVE contribution to the isomerization energy of \emph{ortho}/\emph{meta} benzyne does not exceed 5\%, or $0.21$~kJ/mol. 

\begin{table}[h]
\caption{Final error budget of the calculations of the isomerization energy of \emph{ortho}/\emph{meta} benzyne. All values are given in kJ/mol.}
\label{tab:bugdet-benzyne}
\begin{tabular}{lc}
\hline
 & contribution to $\Delta E$ \\
\hline
Hartree-Fock            & \phantom{$-$}$106.99\pm 0.05$ \\
valence CCSD            & \phantom{0}$-32.32 \pm 0.34$ \\
valence $T_3$           & \phantom{0}$-12.80\pm 0.17$ \\
valence (Q)             & \phantom{00}$-4.92\pm 0.10$ \\
inner-shell correlation & \phantom{00$-$}$2.00\pm 0.10$ \\
scalar relativity       & \phantom{00}$-0.26\pm 0.04$ \\
DBOC                    & \phantom{00$-$}$0.03 \pm 0.10$ \\
ZPVE                    & \phantom{00}$-4.24 \pm 0.21$ \\
\hline
total                   & \phantom{0$-$}$54.52\pm 0.47$ \\
\hline
experiment              & \phantom{00}64.0~[\onlinecite{wenthold98,johnson20}] \\
\hline
other theoretical       & \phantom{000}54.4$^\mathrm{a}$~[\onlinecite{karton09}] \\
                        & \phantom{000}51.5$^\mathrm{b}$~[\onlinecite{ghigo14}] \\
                        & \phantom{000}61.2$^\mathrm{c}$~[\onlinecite{moskaleva99}] \\
                        & \phantom{000}51.0$^\mathrm{d}$~[\onlinecite{lindh95}] \\
\hline
\end{tabular}
\scriptsize\vspace{-0.25cm}
\begin{flushleft}
\hspace{4.0cm} $^\mathrm{a}$ W3.2lite(b) composite method \\
\hspace{4.0cm} $^\mathrm{b}$ CAS(12,12)+PT2/CBS + ZPE$_{\mathrm{CASSCF}}$ \\
\hspace{4.0cm} $^\mathrm{c}$ G2M(rcc,MP2) composite method~\cite{mebel95} \\
\hspace{4.0cm} $^\mathrm{d}$  CASPT2[g1]+aANO:C(5s4p2d)/H(3s2p) basis set
\end{flushleft}
\end{table}

Finally, we consider several minor corrections that do not contribute significantly to the isomerization energy, but are nonetheless required in an accurate study. In the order of importance, we consider first the effect of the inner-shell $1s^2$ orbitals of carbon atoms on the isomerization energy. The inner shell correction was computed as a difference between all-electron and frozen-core CCSD(T) results obtained within the core-valence cc-pwCV$X$Z basis sets~\cite{peterson02}. In this way one obtains $0.64$~kJ/mol, $1.60$~kJ/mol and $1.81$~kJ/mol for $X$=D,T,Q, respectively. Our final estimation, $2.00\pm 0.10$~kJ/mol, is obtained by two-point extrapolation from the $X$=T,Q pair, and the error is estimated in the same fashion as for the valence CCSD contribution. 

The scalar relativistic effects were taken into account using DKH Hamiltonian~\cite{douglas74,hess85,reiher06} as implemented in \textsc{NWChem} program~\cite{jong01}. The relativistic correction was calculated at the all-electron CCSD(T)/cc-pwCV$X$Z level of theory, giving 
$-0.23$, $-0.19$, and $-0.22$ for $X$=D,T,Q, respectively. The final result, $-0.26 \pm 0.02$~kJ/mol, was obtained using the same procedure as for the inner-shell correction. Lastly, the diagonal Born-Oppenheimer correction (DBOC, also known as the adiabatic correction in the literature) was calculated using the CCSD/cc-pVDZ method~\cite{gauss06} employing the \textsc{CFour} program. The result, equal to about $0.03$~kJ/mol, signals that the this effect has a negligible impact on the isomerization energy. While the size of the basis set used is small, and the obtained value is only a rough estimation, it is sufficient for the present purposes. However, we assign large error bars to this quantity, $0.03 \pm 0.10$~kJ/mol.

The results obtained in this section are summarized in Table~\ref{tab:bugdet-benzyne} and compared with other data 
available in the literature. The comparison with the most recent experimental determination~\cite{wenthold98} reveals a substantial 
difference of about $10$~kJ/mol. However, it has to be pointed out that the experimental value was obtained as a 
combination of atomization energies and the error of the final result is difficult to estimate. We find it likely that 
the theoretical value obtained in this work is considerably more accurate which is supported by other theoretical 
results found in the literature. They all tend to cluster around $\Delta E\approx 50-55$~kJ/mol which suggest that the 
experimental value should be revised down.

\subsection{Cope rearrangement in bullvalene molecule}
\label{subsec:application2}

\begin{figure}[ht]
\includegraphics[scale=1.00]{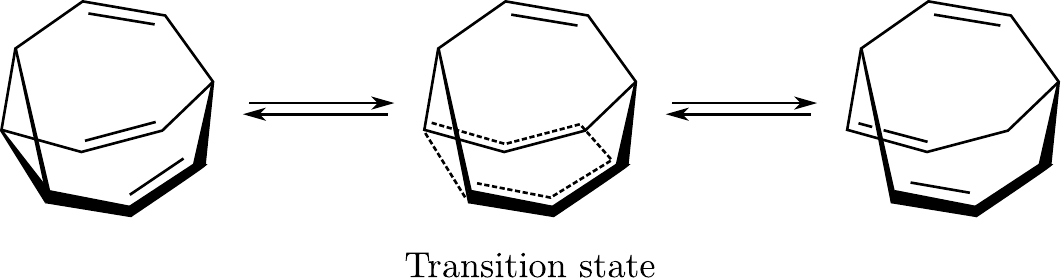}
\caption{\label{fig:bulva-scheme}
Chemical structures illustrating the Cope rearrangement in the bullvalene molecule.
}
\end{figure}

The second system we study in this work in detail is the bullvalene molecule, C$_{10}$H$_{10}$. This molecule attracted considerable attention, because it is a prototypical fluxional molecule that possesses no permanent molecular structure, i.e. the nuclei are constantly in a concerted motion~\cite{ault01}. In bullvalene, this is enabled by the Cope rearrangement, exemplified in Fig.~\ref{fig:bulva-scheme}, that may occur between many equivalent configurations. The initial and final structure are degenerate, but are separated by a reaction barrier. While the bullvalene molecule has been synthesised a long time ago~\cite{doering63,schroder63}, and frequently studied both experimentally and theoretically since then, the height of the barrier is not established unambiguously. The most recent theoretical result of Karton \emph{et al.}~\cite{karton20} differs from the experimental results (obtained by NMR techniques~\cite{moreno92}) by several kJ/mol. This discrepancy is much larger than the reported uncertainties of both calculations and measurements, and hence the theory and experimental data are not consistent at this point. In this section we carry an independent systematic theoretical study of the bullvalene Cope rearrangement barrier height and discuss the possible sources of this inconsistency. In particular, we include corrections due to triple and quadruple excitations calculated with the rank-reduced formalism. These corrections would be extremely costly to compute using the exact CCSDT(Q) method; in fact, we did not manage to accomplish CCSDT(Q) calculations even with the smallest cc-pVDZ basis.

The electronic contribution to the reaction barrier height is denoted by the symbol $\Delta E^\ddag$. For the purposes of direct comparison with the experimental data, we additionally need to calculate the Gibbs free energy barrier heights at the temperature $T=298$~K. This quantity is denoted by $\Delta G_{298}^\ddag$ and includes, besides $\Delta E^\ddag$, the zero-point vibrational energy (ZPVE) and enthalpic/entropic temperature corrections, as detailed below.

The molecular geometries of the bullvalene equilibrium structure and Cope rearrangement transition state were optimized at the B3LYP-D3/pc-2 level of theory~\cite{grimme10,grimme11,jensen01,jensen02} using \textsc{NWChem} package. The obtained structures were verified to represent the equilibrium structure (real harmonic frequencies) and first-order transition state (one imaginary frequency). The Cartesian geometries of both structures are given in Supporting Information. The barrier height $\Delta E^\ddag$ is split into several components calculated at different levels of theory, and a composite scheme is used to assemble the best theoretical estimate. Because the bullvalene molecule is roughly twice as large as the systems considered in Sec.~\ref{subsec:application1}, the composite method applied here is less rigorous in nature. In particular, we do not assign uncertainties to individual contributions to the barrier height; instead, we attach a global error estimate only to the final result.

\begin{figure}[ht]
\includegraphics[scale=0.75]{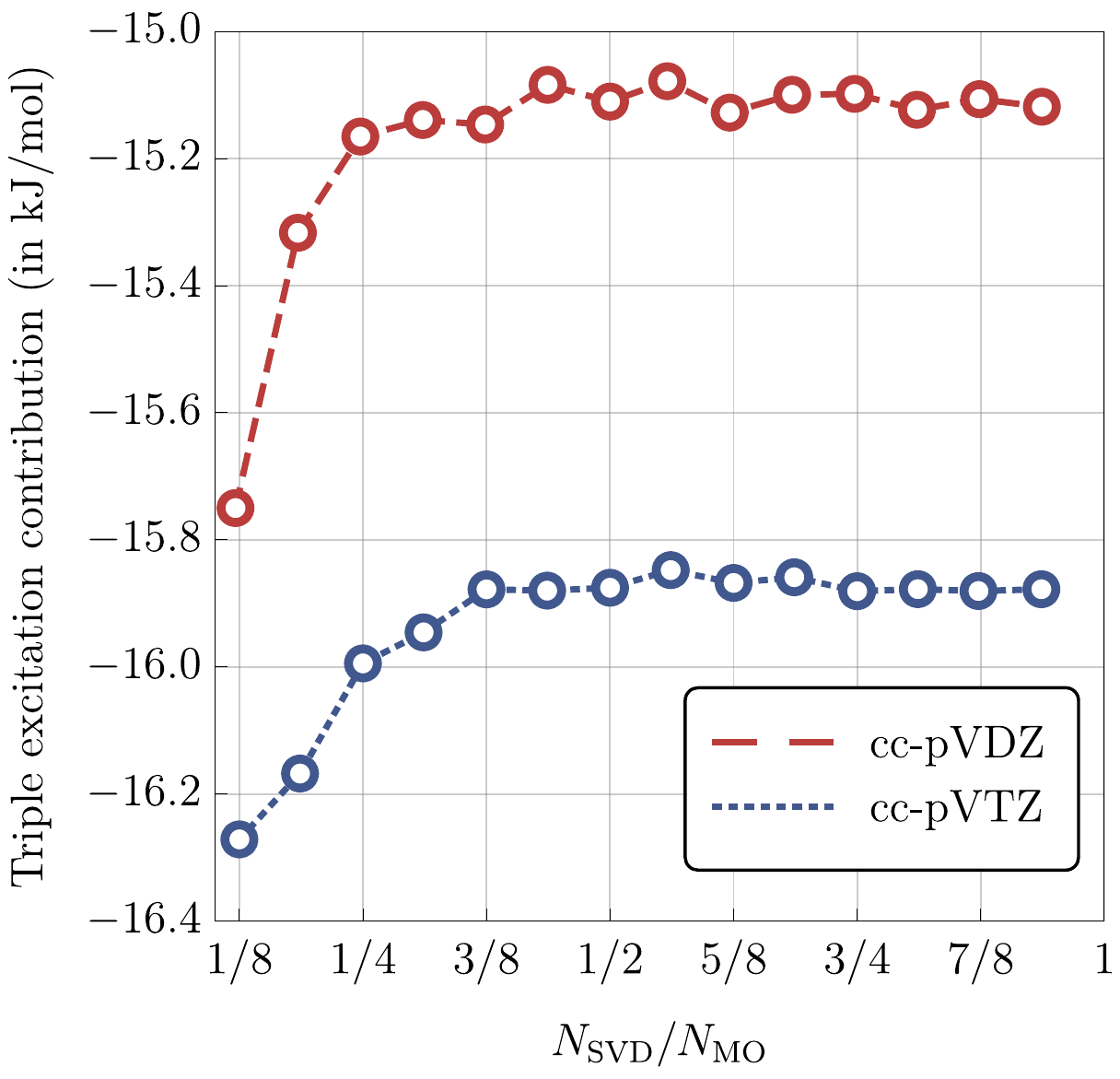}
\caption{\label{fig:t-bulva}
Triple excitation contribution to the Cope rearrangement barrier height ($\Delta E^\ddag$) in bullvalene molecule calculated using the SVD-CCSDT+ method (cc-pVDZ and cc-pVTZ basis sets) as a function of the $\nsvd$ parameter. The symbol $N_{\mathrm{MO}}$ denotes the total number of orbitals in the system.
}
\end{figure}

First, we consider the Hartree-Fock contribution to the barrier height which was calculated using the cc-pV$X$Z basis sets with $X$=T,Q,5. The exponential extrapolation (\ref{extra_exp}) from these three basis sets leads to the result $108.97$~kJ/mol. This differs by less than $0.1$~kJ/mol from the result obtained within the cc-pV5Z basis, showing that the error of the Hartree-Fock component of $\Delta E^\ddag$ is negligible. The second contribution to $\Delta E^\ddag$ was calculated using the CCSD method, giving $-29.20$, $-28.39$, and $-27.50$~kJ/mol with cc-pV$X$Z, $X$=D,T,Q, basis sets, respectively. To further reduce the basis set incompleteness error we apply the two-point extrapolation formula (\ref{extra_riemann}) resulting in the final CCSD contribution of $-26.68$~kJ/mol.

Next, we consider the contribution of triple excitations to the barrier height. It was computed using the SVD-CCSDT+ method within the cc-pVDZ and cc-pVTZ basis sets. Similarly as in the previous section, we do not split the effect of triple excitations into (T) and post-(T) components, because we did not manage to calculate the (T) correction within a larger (cc-pVQZ) basis set due to excessive time requirements. Note that the SVD-CCSDT+ calculations within the cc-pVTZ basis involve 50 correlated electrons and 440 atomic orbitals which vastly exceeds the capabilities of the available CCSDT implementations. In Fig.~\ref{fig:t-bulva} we present triple excitation contribution to the barrier height as a function of the $\nsvd$ parameter. The results saturate fast with respect to the value of $\nsvd$, and for $\nsvd=\half N_{\mathrm{MO}}$ they are essentially converged. Beyond this point minor fluctuations at the level of ca. $0.05$~kJ/mol and $0.02$~kJ/mol are observed, but this is completely negligible in comparison with other sources of error. Using the results obtained with $\nsvd=N_{\mathrm{MO}}$, we obtain the contributions of triple excitations equal to $-15.23$ and $-15.87$~kJ/mol in the cc-pVDZ and cc-pVTZ basis sets, respectively. The final result, $-16.26$~kJ/mol, is obtained using two-point extrapolation formula, Eq.~(\ref{extra_riemann}).

\begin{figure}[ht]
\includegraphics[scale=0.75]{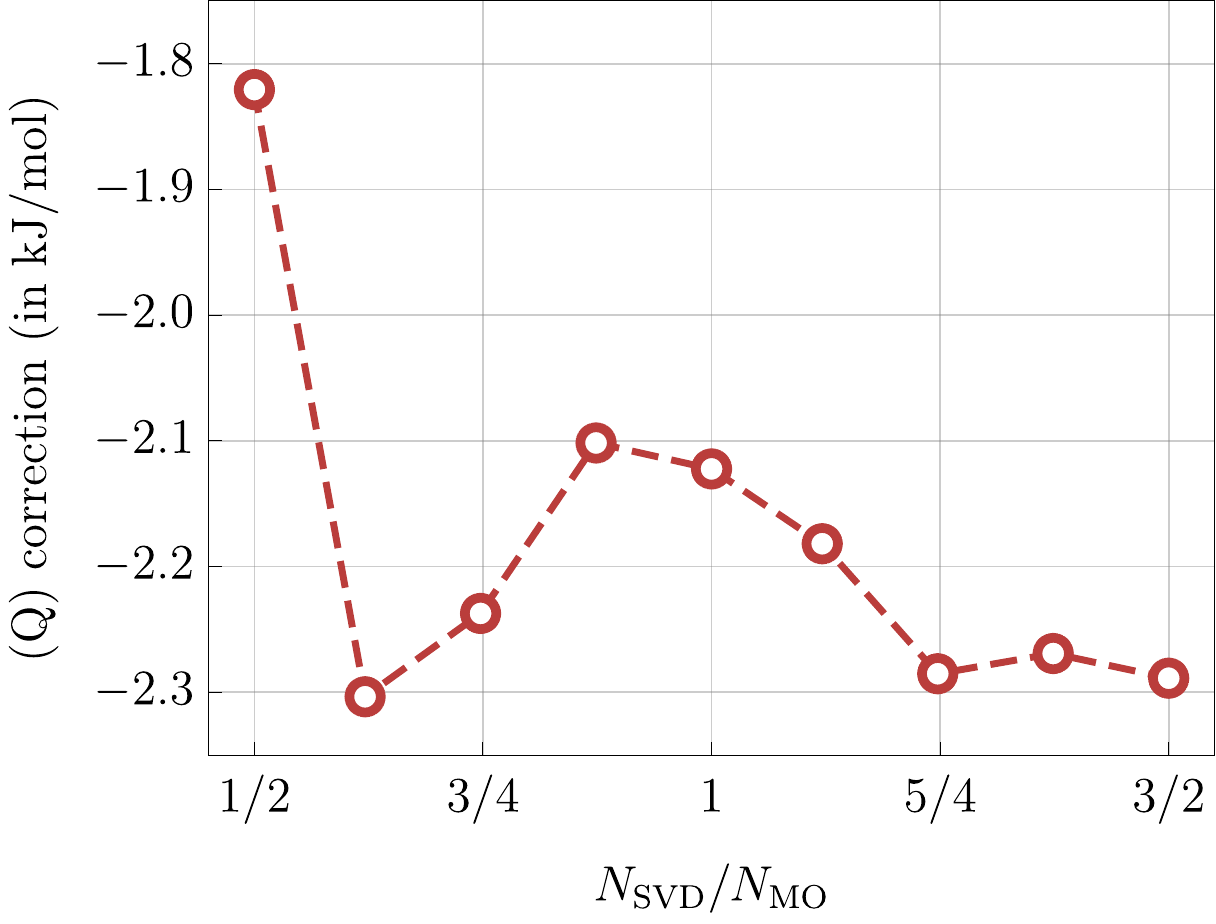}
\caption{\label{fig:q-bulva}
Quadruple excitation contribution to the Cope rearrangement barrier height ($\Delta E^\ddag$) in bullvalene molecule calculated using the $\mathcal{L}_{\rm (Q)}$ functional (cc-pVDZ basis set) as a function of the $\nsvd$ parameter. For each value of $\nsvd$, the parameter $\nqua$ is set to $\frac{2}{3}\nsvd$. The symbol $N_{\mathrm{MO}}$ denotes the total number of orbitals in the system.
}
\end{figure}

The quadruple excitation contribution to $\Delta E^\ddag$ was calculated using the $\mathcal{L}_{\rm (Q)}$ functional and the cc-pVDZ basis set. In Fig.~\ref{fig:q-bulva} we present the (Q) correction as a function of the $\nsvd$ parameter and with the recommended $\nqua=\frac{2}{3}\nsvd$. One can see that beyond $\nsvd=\frac{5}{4}N_{\mathrm{MO}}$ the results are essentially stable with respect to this parameter. The variations are within $0.01-0.02$~kJ/mol and hence are negligible from the present point of view. Therefore, we take the value obtained with $\nsvd=\frac{3}{2}N_{\mathrm{MO}}$, namely $-2.29$~kJ/mol, as the final contribution of quadruple excitations to $\Delta E^\ddag$. The computations of the (Q) correction using the quadratic functional for the bullvalene molecule with $\nsvd=\frac{5}{4}N_{\mathrm{MO}}$ and $\frac{2}{3}\nsvd$ (cc-pVDZ basis set) take about 3 days on 14 cores of AMD Opteron$^{\mathrm{TM}}$ Processor 6174.

The last contributions to $\Delta E^\ddag$ are due to the inner-shell correlation and relativistic effects (scalar DKH Hamiltonian). They were both calculated using all-electron CCSD method within cc-pwCVTZ basis set supplemented by (T) correction obtained within cc-pwCVDZ basis. No extrapolation towards the complete basis set was performed. This brings contributions to $\Delta E^\ddag$ equal
$1.33$ and $-0.21$~kJ/mol due to the aforementioned two effects. We also estimated the DBOC component of $\Delta E^\ddag$ (CCSD/cc-pVDZ level of theory) and found it to be negligible ($<0.1$~kJ/mol) within the present accuracy standards.

\begin{table}[h]
\caption{Summary of the calculations of the Gibbs free energy barrier height for the Cope rearrangement in the bullvalene molecule. All values are given in kJ/mol.}
\label{tab:bugdet-bullvalene}
\begin{tabular}{lc}
\hline
 & contribution to $\Delta G_{298}^\ddag$ \\
\hline
Hartree-Fock            & \phantom{$-$}$108.97$ \\
valence CCSD            & \phantom{0}$-26.68$ \\
valence $T_3$           & \phantom{0}$-16.26$ \\
valence (Q)             & \phantom{00}$-2.29$ \\
inner-shell correlation & \phantom{00$-$}$1.33$ \\
scalar relativity       & \phantom{00}$-0.21$ \\
ZPVE                    & \phantom{00}$-4.36$ \\
thermal correction      & \phantom{00$-$}$0.44$ \\
\hline
total                   & \phantom{$-$0}60.94 \\
\hline
\end{tabular}
\end{table}

Finally, ZPVE contribution to the barrier height, as well as thermal corrections, were calculated at the same level of theory as the geometry optimization (B3LYP-D3/pc-2). The raw value of ZPVE was additionally scaled by the empirical factor $f=0.9678$, as recommended in Ref.~\onlinecite{heine19}, to take the anharmonic effects into account, giving $-4.36$~kJ/mol. Thermal corrections were calculated within the rigid rotor/harmonic oscillator approximations without frequencies scaling. The thermal enthalpic and entropic contributions to the Gibbs free energy barrier height for $T=298$~K are $-0.59$ and $1.03$~kJ/mol, respectively, and hence the total finite-temperature correction is just $0.44$~kJ/mol.

The final results of the calculations of the Gibbs free energy barrier height for the Cope rearrangement in the bullvalene molecule are summarized in Table~\ref{tab:bugdet-bullvalene}. The total $\Delta G_{298}^\ddag$ determined by us equals to $60.94$~kJ/mol. In order to roughly estimate the error of this result we note that there are two major sources of uncertainty: valence CCSD and ZPVE contributions. They can both lead to errors of the order of $0.5$~kJ/mol. The remaining contributions to $\Delta G_{298}^\ddag$ are expected to be accurate to within $0.1-0.2$~kJ/mol. All in all, the value determined by us, $\Delta G_{298}^\ddag=60.94$~kJ/mol, has uncertainty of around $1$~kJ/mol. This result is in reasonable agreement with the most recent theoretical result of Karton~\cite{karton20}, $62.21$~kJ/mol, but in a disagreement with older calculations based on lower levels of theory which give results within $35-55$~kJ/mol range~\cite{hrovat05,brown09,greve11,khojandi20}. More strikingly, our result is in a disagreement with the experimental data of Moreno \emph{at al.}~\cite{moreno92} who obtained $\Delta G_{298}^\ddag=54.8\pm 0.8$~kJ/mol from gas-phase NMR measurements. Such a large difference of about $6$~kJ/mol is unlikely to be caused by an error in the theoretical protocol adopted by us. Therefore, we believe that the experimental data for this system should be reevaluated and a new measurement may help to resolve the persisting discrepancy between state-of-the-art theory and experimental results.

\section{Conclusions and future work}
\label{sec:conclusion}

In this work we have extended the rank-reduced coupled-cluster formalism to the calculation of non-iterative energy corrections due to quadruple excitations. The focus of the present work has been concentrated on the CCSDT(Q) method, which has become \emph{de facto} standard in high-accuracy \emph{ab initio} quantum chemistry, and can be viewed as the ``platinum standard'' of the field. The proposed formalism consist of two major novel components. The first is the application of the Tucker format to compress the quadruple excitation amplitudes and eliminate the full rank $t_{ijkl}^{abcd}$ tensor entirely from the computational procedure. The second is the introduction of a modified functional for evaluation of the (Q) correction. This functional is rigorously equivalent to the standard (Q) formalism when the exact CC amplitudes are used. However, due to the fact that the new functional is stationary with respect to the amplitudes, it is less susceptible to errors resulting from the aforementioned compression. We show, both theoretically and numerically, that the computational cost of the proposed method scales as the seventh power of the system size. Using reference results for a set of small molecules, the method is calibrated to deliver accuracy of a few percent in relative energies. To illustrate the potential of the theory we calculate the isomerization energy of \emph{ortho}/\emph{meta} benzyne (C$_6$H$_4$) and the barrier height for the Cope rearrangement in bullvalene (C$_{10}$H$_{10}$). In both cases we show that the proposed formalism considerably increases the range of applicability of the CC theory with non-iterative energy corrections due to quadruple excitations.

The present work is a starting point for a rank-reduced treatment of other quantum chemistry methods involving quadruple excitations. Indeed, the quadruple excitation subspace obtained by the HOOI procedure, Sec.~\ref{subsec:t4hooi}, can be used also in more advanced (both iterative and non-iterative) CC models involving the $T_4$ operator. This includes even the complete CCSDTQ method. In fact, our preliminary study showed that the $N^7$ scaling can be achieved at the CCSDTQ level if both the triple and quadruple excitation amplitudes are compressed using the Tucker format. However, to exploit this advantage an efficient implementation is required to minimize the prefactor, and the accuracy of the resulting method must be thoroughly tested and calibrated.

Another important extension is generalization of the rank-reduced coupled-cluster formalism to the 
open-shell situations. This direction is especially important for applications in \emph{ab initio} 
thermochemistry, where calculation of atomization energies is an important problem. A 
straightforward way to handle the open-shell systems is offered by the spin-unrestricted 
coupled-cluster theory, but this approach leads to the spin-contamination of the electronic wavefunction, as is well-documented in the literature~\cite{stanton94,krylov00,kitsaras21}. While the issue of spin-contamination may not be severe in many 
applications, a more pressing problem is the need to handle numerous spin cases of the triply- 
and, especially, quadruply-excited configurations. In the spin-unrestricted formalism each spin 
case has to be decomposed separately, leading to a significant increase in the computational costs. 
As a result, a more robust and advanced~\cite{rittby88,knowles93,jeziorski95,crawford97,szalay97,jankowski99,heckert06} approach to the spin-adaptation in open-shell systems may be required which will be considered in future works.

\begin{acknowledgement}
I would like to thank Dr. A. Tucholska for fruitful discussions, and for reading and commenting on the manuscript. The author is grateful to Dr. D. Matthews for his help in effective usage of the \textsc{TBlis} library.
This work was supported by the National Science Center, Poland, through the project No. 2017/27/B/ST4/02739.  Computations presented in this research were carried out with the support of the  Interdisciplinary Center for Mathematical and Computational Modeling (ICM) at the University of Warsaw, grant numbers G86-1021 and G88-1217. 
\end{acknowledgement}

\begin{suppinfo}
The following file is available free of charge via the Internet at \texttt{http://pubs.acs.org}:
\begin{itemize}
  \item {\tt quad-supp.pdf}: additional derivations and technical details of the proposed computational procedure;
  \item {\tt geometries.tar}: optimized molecular geometries from Sec.~\ref{subsec:param} in the \texttt{*.xyz} format.
\end{itemize}

\end{suppinfo}

\bibliography{svd_quad}

\end{document}